\makeatletter
\let\orig@ProvidesClass\ProvidesClass
\def\ProvidesClass#1{\orig@ProvidesClass{JFM-FLM_Au}}
\def\thepage{\arabic{page}}
\@namedef{ver@breakurl.sty}{2013/04/10}
\@namedef{ver@relsize.sty}{2013/03/29}
\newcommand{\relsize}[1]{}
\let\orig@definecounter\@definecounter
\def\@definecounter#1{%
  \expandafter\let\csname the#1\endcsname\relax
  \orig@definecounter{#1}%
}
\newskip\abovecaptionskip
\newskip\belowcaptionskip
\long\def\@makecaption#1#2{%
  \vskip\abovecaptionskip
  \sbox\@tempboxa{#1: #2}%
  \ifdim \wd\@tempboxa >\hsize
    #1: #2\par
  \else
    \global \@minipagefalse
    \hb@xt@\hsize{\hfil\box\@tempboxa\hfil}%
  \fi
  \vskip\belowcaptionskip}
\newdimen\paperwidth
\newdimen\paperheight
\makeatother

\RequirePackage{array}
\RequirePackage[dvipsnames]{xcolor}
\documentclass[paper]{JFM-FLM_Au}
\makeatletter
\gdef\@underjournal{%
  \vbox to 5.5\p@{\noindent
    \parbox[t]{4.5in}{\normalfont\indexsize{\itshape Under consideration
      for publication in J.\ Fluid\ Mech.}}%
  \vss}%
}
\def\ps@titlepage{\leftskip\z@\let\@mkboth\@gobbletwo\vfuzz=5\p@
  \def\@oddhead{\vbox to \headheight{\vspace*{-4pt}\hbox to \textwidth{\@j@urnal \hfill}\par\vskip4pt%
    \hbox to \textwidth{{\fboxsep0pt\fbox{\parbox{\dimexpr\textwidth-2\fboxrule\relax}{\par\vskip2.7pc\centerline{Banner appropriate to article type will appear here in typeset article}\par\vskip2.7pc}}}}\vss}}%
  \def\@evenhead{\@oddhead}%
  \def\@oddfoot{\oddabsfooterflag}%
  \def\@evenfoot{\evenabsfooterflag}%
  \def\sectionmark##1{}%
  \def\subsectionmark##1{}%
}
\def\pagelimitfooter{\bgroup%
\ifnum\c@page=4\relax%
\evenfooterflagdefns{Focus on Fluids articles must not exceed this page length}%
\else%
\ifnum\c@page=10\relax%
\evenfooterflagdefns{Rapids articles must not exceed this page length}%
\else%
{\hbox to \textwidth{{\cppagefont\ifx\@volume\undefined\else\textbf{\@volume}\fi\ \ifx\@issue\undefined\ifpaper A\else\ifrapid R\else\ifpersp P\else\iffof F\else X\fi\fi\fi\fi1-\else\ifpaper A\else\ifrapid R\else\ifpersp P\else\iffof F\else X\fi\fi\fi\fi\@issue-\fi\thepage}\hfill}}%
\fi\fi%
\egroup}%
\def\evenfooterflagdefns#1{}
\def\oddfooterflagdefns#1{}
\def\evenabsfooterflag{}
\def\oddabsfooterflag{}
\AtBeginDocument{%
  \pdfpagewidth=174mm
  \pdfpageheight=247mm
}
\makeatother

\usepackage{tabularx}
\usepackage{makecell}
\usepackage{adjustbox}

\lefttitle{A. Chaudhary, M. Chamecki and M. Heisel}
\righttitle{Journal of Fluid Mechanics}

\title{Turbulent velocity statistics in convective boundary layers}

\author{Aditya Chaudhary\aff{1}, Marcelo Chamecki\aff{2} \and Michael Heisel\aff{1}}

\affiliation{\aff{1}School of Civil Engineering, The University of Sydney, NSW 2008, Australia
\aff{2}Department of Atmospheric and Oceanic Sciences, University of California, Los Angeles, CA, USA}

\corresau{Michael Heisel, \email{michael.heisel@sydney.edu.au}}
\begin{document}
\maketitle
\begin{abstract}
Large-eddy simulations (LES) of the convective atmospheric boundary layer (CBL) are used to investigate velocity variance statistics in turbulent sheared convection. The thirteen simulations span conditions from weak convection with longitudinal rolls to free penetrative convection with no mean wind. In the presence of thermal rolls, the near-surface streamwise variance profile exhibits an approximately logarithmic decay with a slope proportional to the friction velocity, $u_*$, and an added constant proportional to the convective velocity, $w_\star$. After the structural transition to convective cells, both the streamwise and spanwise variances follow the empirical stability and height dependencies of Wilson (\textit{Boundary-Layer Meteorol.}, 2008, vol. 129, pp. 353--369). The near-surface vertical (wall-normal) velocity variance is well described by Monin--Obukhov similarity with improved agreement using local-in-height scaling. Within the mixed layer, the peak vertical variance initially decreases relative to $w_\star$ as shear diminishes, before increasing and shifting downward as the flow approaches free convection. In contrast, the total turbulent energy decreases relative to $w_\star$ across the parameter space. The spanwise velocity variance represents the dominant turbulent component under weak convection and exceeds the streamwise variance throughout most of the convective regime. These statistical trends are attributed to the role of pressure, evaluated through variance budget contributions. Large-scale buoyancy production in the mixed layer is transported downward by pressure and redistributed preferentially to the spanwise component, coinciding with the presence of longitudinal roll motions. The weakening of this pressure-transport pathway for convective cells coincides with greater retention of vertical variance within the mixed layer.
\end{abstract}
\vskip14.5pt\hrule\vskip 18pt

\section{Introduction}
\label{sec:intro}

Sheared convective boundary layers represent a fundamental class of fluid flows with widespread relevance across engineering and the natural sciences. These environments are characterised by the complex interaction between wall-bounded shear turbulence and positive surface buoyancy \citep{Moeng1994, Khanna1998, Salesky2017, BouZeid2018, ChavezMedina2025}, with applications ranging from predicting heat transfer in industrial thermal management systems \citep{Blass2020, Wang2021} to understanding land-atmosphere interactions \citep{Albertson1999, Stoll2020}. The convective atmospheric boundary layer (CBL) serves as a prominent and widely studied example of this physical system within the atmospheric sciences, controlling the vertical transport of momentum, energy, moisture, and trace chemicals between the Earth's surface and the free troposphere \citep{Brutsaert1982, Stull1988, Garratt1992, Wyngaard2010, Dewani2023}.

A notable characteristic of these environments is the large-scale structural organisation of the turbulent motions. As the relative influence of buoyancy and shear changes, the flow undergoes a transition from highly-organised longitudinal rolls aligned with the mean wind to buoyancy-driven polygonal cells \citep{Sykes1989, Khanna1998, Weckwerth1999, Salesky2017}. The topological shift from rolls to cells occurs for all sheared convection configurations, including wall-sheared Rayleigh--B{\'e}nard convection and various internal flows \citep{Pandey2018, Blass2020, Wang2020}. This universality suggests that these structures and their dynamical influence have broad relevance to fundamental fluid mechanics across applications. Despite the well-documented visual evolution of these large-scale structures, the turbulent second-order statistics associated with these structural transitions are less well understood and agreed upon \citep{Salesky2017, Zhou2019}. Further, the exact physical mechanisms by which turbulent kinetic energy (TKE) is produced, redistributed among the velocity components, and dissipated as the flow reorganises remain an active area of research \citep{Ding2018, BouZeid2018}. 

Within the surface layer of the CBL, variance statistics and profiles are traditionally evaluated within the framework of Monin--Obukhov similarity theory (MOST), which does not explicitly account for the large-scale structural organisation. MOST postulates that these near-surface turbulent statistics follow a universal function of the surface friction velocity $u_*$ and a local stability parameter $\zeta = z/L$ \citep{Foken2006}, where $z$ indicates the height (or wall-normal distance) above the surface and $L$ is the Obukhov length:
\begin{equation}\label{eq:1}
L = -\frac{u_*^3 \theta_0}{\kappa g Q}.
\end{equation}
Here, $\kappa \approx 0.4$ is the von K{\'a}rm{\'a}n constant, $g$ is gravitational acceleration, $\theta_0$ is the reference potential temperature, $Q$ is the surface kinematic heat flux, and $L$ is negative for unstable convection with positive $Q$. The global importance of shear versus buoyancy across the entire boundary layer is typically represented by the bulk stability parameter $-z_i/L$, where $z_i$ is the overall CBL depth. In this framework, $-z_i/L \to 0$ corresponds to a neutrally-stratified boundary layer with no net buoyancy, while $-z_i/L \to \infty$ corresponds to the absolute free-convection limit where mean wind shear is entirely absent. The conventional equation forms derived from MOST to model the normalised variances of the fluctuating streamwise ($u'$), spanwise ($v'$), and vertical (wall-normal, $w'$) velocity components are typically empirical power-law functions of $z/L$ or $z_i/L$. While historically foundational, this framework remains strictly constrained to the aforementioned surface layer (i.e. logarithmic region), comprising the lowest $10\%$ of the CBL in the high-Reynolds-number limit \citep{Garratt1992, Yaglom1994}. Accurately mapping the behaviour of the turbulence statistics across the parameter space of sheared convection therefore requires consideration of the neutral and free-convection limits, in addition to the constraints imposed by the outer layer farther from the surface.

\citet{Deardorff1970} used dimensional analysis to predict the scaling of turbulence in buoyancy-dominated conditions. The scales $u_*$ and $z$ associated with wall-bounded shear become irrelevant both in the global limit of free convection and far from the surface in the convective mixed layer. Instead, large-scale turbulent behaviour is controlled by the surface kinematic heat flux and the CBL depth, which combine to form the convective velocity scale $w_\star = ( g Q z_i / \theta_0 )^{1/3}$ \citep{Deardorff1970}. The balance of the shear and buoyancy velocity scales is directly related to the bulk stability parameter as $u_*/w_\star = \kappa^{1/3} (-z_i/L)^{-1/3}$, providing a direct connection between  $-z_i/L$ and the transition in scaling of turbulent velocities in these flows.

Conversely, in the neutral limit $-z_i/L \to 0$, turbulence is governed by the aforementioned $u_*$ and $z$ scales. Variance statistics are well-predicted by the attached eddy hypothesis in which eddies have size confined by the distance from the wall \citep{Townsend1976}. The vertical velocity variance is approximately constant in the near-surface logarithmic region, while the horizontal variances exhibit a logarithmic decay with height:
\begin{equation}\label{eq:2}
\frac{\langle u'^2 \rangle}{u_*^2} = B_u - A_u \log{\left( z / z_i \right)}.
\end{equation}
Here, $A_u$ and $B_u$ are constant coefficients and angle brackets indicate ensemble averaging. Extensive experimental and numerical evidence from wall-bounded shear flows has since supported these neutral-limit predictions \citep[see, e.g.,][]{Marusic2019,Baars2020,Yang2020,Qin2025}, establishing baseline structural profiles for purely shear-driven turbulence. As seen in table~\ref{tab:vertical}, the constant value for the vertical variance is in the range $B_w \approx $ 1.45 to 1.9 \citep{Panofsky1984,Perry1990,Heisel2026}. Values for the constants $A_u$ and $B_u$ in \eqref{eq:2} are given in table~\ref{tab:streamwise}, noting $B_u$ is dependent on the flow configuration. Both variances are known to have corrections for finite Reynolds numbers \citep{Perry1987,Bernardini2014,Pirozzoli2024,Heisel2026}, but only asymptotically high Reynolds number CBLs are considered here following the traditional assumption for atmospheric flows.

Historical predictions for the vertical velocity variance in the CBL are consistent with the constraints discussed above. The most common functional form shown in table~\ref{tab:vertical} for the convective surface layer spans from the constant neutral baseline to convective scaling, and traces the transition between them via MOST. In MOST, the convective velocity in the free-convection limit is a variation of $w_\star$ defined using $z$ rather than $z_i$ due to the definition of $\zeta$. This limit is known as local free convection owing to the dependence on the local-in-height stability $\zeta$ as opposed to global convection based on $z_i/L$ \citep{Wyngaard1971}. Observations of $\langle w'^2 \rangle$ across a diverse range of simulations and atmospheric field measurements are well described by this functional form \citep[e.g.,][]{Kaimal1976, Panofsky1977, Khanna1997, Wilson2008, Maronga2017}. For the bulk of the CBL above the surface layer, universal functions for $\langle w'^2 \rangle$ have been proposed using the global convective velocity $w_\star$ and the relative height $z/z_i$ \citep{Lenschow1980}. Similar to the surface layer, measurements and simulations have been shown to fit this functional form with some variability in the coefficients and exponents \citep[e.g.,][]{Caughey1979, Sorbjan1988, Lenschow2012, Zhou2019, Dewani2023}. Other formulations and conditions for $\langle w'^2 \rangle$ have been proposed in the literature -- such as those incorporating shear-buoyancy superposition \citep{Zilitinkevich1994, Zilitinkevich2006}, local buoyancy-shear scaling \citep{Nieuwstadt1984, Sorbjan1986}, or dependence on anisotropy \citep{Stiperski2023} -- but these are not evaluated in the present analysis, aside from local scaling in the surface layer.

\begin{table}
\centering
\footnotesize
\setlength{\tabcolsep}{4pt}
\renewcommand{\arraystretch}{1.2}
\renewcommand{\tabularxcolumn}[1]{p{#1}}

\begin{tabularx}{\textwidth}{
    >{\raggedright\arraybackslash}X
    >{\raggedright\arraybackslash}p{0.27\textwidth}
    >{\raggedright\arraybackslash}p{0.16\textwidth}
    >{\raggedright\arraybackslash}X
}
Condition & Formulation & Empirical constants & Sources \\
\hline

\makecell[tl]{Neutral surface (log) layer}
&
$\langle w'^2\rangle/u_*^2 = B_w$
&
$B_w \approx 1.45\text{--}1.9$
&
\citet{Townsend1976};
\citet{Panofsky1984};
\citet{Perry1990};
\citet{Kunkel2006};
\citet{Heisel2026}
\\

\makecell[tl]{Convective surface layer}
&
$\langle w'^2\rangle/u_*^2
=
B_{w_1}^2(1-A_{w_1}\zeta)^{2/3}$
&
\makecell[tl]{
$A_{w_1}\approx3.0\text{--}9.5$\\
$B_{w_1}\approx0.8\text{--}1.3$\\
}
&
\citet{Panofsky1977};
\citet{Wilson2008};
\citet{Maronga2017};
\citet{Kaimal1994};
\citet{Khanna1997}
\\

\makecell[tl]{Local free-convection \\
surface layer ($-\zeta \gg 1$)}
&
$\langle w'^2 \rangle/u_*^2 = A_{w_2}^2(-\zeta)^{2/3}$
&
$A_{w_2} \approx 1.9$
&
\citet{Priestley1954}; 
\citet{Wyngaard1971}
\\

\makecell[tl]{Convective mixed layer}
&
$\langle w'^2\rangle/w_\star^2
=
B_{w_3}(z/z_i)^{2/3}
\left(1-A_{w_3}\,z/z_i\right)^p$
&
\makecell[tl]{
$A_{w_3}\approx0.8\text{--}1.0$\\
$B_{w_3}\approx1.17\text{--}1.8$\\
$p\in\{2/3,\,2\}$
}
&
\citet{Lenschow1980};
\citet{Sorbjan1988};
\citet{Zhou2019};
\citet{Dewani2023}
\\

\hline
\end{tabularx}

\caption{Common formulations for predicting the vertical velocity variance $\langle w'^2 \rangle$ across different conditions within convective boundary layers (CBLs), noting some formulas are derived from definitions for the standard deviation.}

\label{tab:vertical}
\end{table}

Unlike the vertical variance, early empirical models for the streamwise variance $\langle u'^2 \rangle$ in the CBL were inconsistent with the neutral limit in \eqref{eq:2}. The first predictions for near-neutral conditions considered a constant $\langle u'^2 \rangle/u_*^2 \approx B_{u_2}$ with no height dependence \citep{Lumley1964}. \citet{Panofsky1977} later added a $z_i/L$ correction to $B_{u_2}$ to account for a wider range of convection. This formulation follows the assumption that horizontal velocity fluctuations in the surface layer are largely governed by large-scale, buoyancy-driven convective structures that extend throughout the mixed layer, rendering the boundary-layer depth $z_i$ the relevant scaling parameter rather than $\zeta$ (MOST). More recently, \citet{Wilson2008} incorporated a power-law height dependence with the \citet{Panofsky1977} relation to better fit field observations. Further efforts to address the limitations of MOST in parameterising $\langle u'^2 \rangle$ include introducing a horizontal length scale for multipoint Monin--Obukhov similarity \citep{TongNguyen2015} and predicting variances based on anisotropy of the Reynolds-stress tensor \citep{Stiperski2023}. However, these frameworks do not explicitly account for the logarithmic profile and presence of attached eddies in the neutral limit.

\citet{Banerjee2015} first attempted to reconcile $\langle u'^2 \rangle$ in convective conditions with \eqref{eq:2} using a spectral budget method, yielding the same logarithmic equation but with the coefficients depending on local stability $\zeta$. In an alternate approach, \citet{Yang2020} relaxed Townsend's \citeyearpar{Townsend1976} assumption of an attached eddy population to relate $\langle u'^2 \rangle$ to the mean flow deficit, with the model reducing to the form in \eqref{eq:2} under neutral conditions when the mean velocity $U(z)$ follows a logarithmic profile. Both methods represent structural modifications to the attached eddies under weak to moderate convection, but have received limited corroboration across a wide range of conditions.

Further aloft in the convective mixed layer where mechanical shear is negligible,  $\langle u'^2 \rangle$ is characterised by the convective velocity scale $w_\star$. While $\langle u'^2 \rangle$ exhibits systematic height and stability dependence in the lower mixed layer due to the interaction of near-surface shear and convective eddies \citep{Khanna1997, Khanna1998}, it has historically been approximated in the outer mixed layer as a vertically uniform profile \citep{Panofsky1977, Caughey1979}. The approximately constant variance with respect to $w_\star$ reflects the dominant role of non-local, boundary-layer-spanning convective structures that carry outer-layer velocity scales down to the surface \citep{Zilitinkevich2006}. 

\begin{table}
\centering
\footnotesize
\setlength{\tabcolsep}{4pt}
\renewcommand{\arraystretch}{1.2}
\begin{tabularx}{\textwidth}{ 
    >{\hsize=0.55\hsize\raggedright\arraybackslash}X 
    >{\hsize=1.40\hsize\raggedright\arraybackslash}X 
    >{\hsize=0.80\hsize\raggedright\arraybackslash}X 
    >{\hsize=1.25\hsize\raggedright\arraybackslash}X} 
Condition & Formulation & Empirical constants & Sources \\
\hline

Neutral surface (log) layer
&
$\langle u'^2 \rangle/u_*^2 = B_u-A_u\log(z/z_i)$
&
$A_u \approx 0.91\text{--}1.33$, $B_u$ flow dependent
&
\citet{Townsend1976}; \citet{Perry1987}; \citet{Marusic2013}; \citet{Bernardini2014}; \citet{Banerjee2015}; \citet{Baars2020}; \citet{Huang2022}
\\

Convective surface layer
&
$\langle u'^2 \rangle/u_*^2 = \left[ 12 + 0.5(-z_i/L) \right]^{2/3}$
&
---
&
\citet{Panofsky1977}
\\

&
\makecell[l]{$\displaystyle \frac{\langle u'^2 \rangle}{u_*^2} = \left[ 4+0.6 \left(\frac{-z_i}{L}\right)^{2/3} \right]$ \\ $\displaystyle \qquad \times \left[ 1-\left(\frac{z}{z_i}\right)^{0.38} \right]$}
&
---
&
\citet{Wilson2008}
\\

&
$\langle u'^2 \rangle/u_*^2 = B_{u}-A_{u}\log(z/z_i)$
&
$A_{u} = f(\zeta)$, $B_{u} = f(\zeta)$
&
\citet{Banerjee2015}
\\

&
$\langle u'^2 \rangle/u_*^2 = A_{u_2}\left[U_\infty/u_*-U/u_*\right]+B_{u_2}$
&
$A_{u_2} = f(-z_i/L)$, $B_{u_2} = f(-z_i/L)$
&
\citet{Yang2020}
\\

Convective mixed layer
&
$\langle u'^2 \rangle/w_\star^2 = f(z/z_i, -z_i/L) \approx B_{u_3}$
&
$B_{u_3} \approx 0.35$
&
\citet{Caughey1979}; \citet{Khanna1997}; \citet{Khanna1998}
\\

\hline
\end{tabularx}
\caption{Same as table~\ref{tab:vertical}, except the formulations are for the streamwise velocity variance $\langle u'^2 \rangle$. Here, $U_\infty$ denotes the outer or reference mean streamwise velocity.}
\label{tab:streamwise}
\end{table}

For the spanwise (cross-wind) variance, $\langle v'^2 \rangle$, the logarithmic spatial decay evaluated in \eqref{eq:2} is consistent with predictions and observations under neutral conditions \citep{Townsend1976, Lee2015, Zimmerman2019, Baidya2021}. Within the convective surface layer, proposed formulations for $\langle v'^2 \rangle$ generally follow the same empirical corrections applied to the streamwise component: the variance is evaluated as a constant relative to $u_*$ in the neutral limit of MOST \citep{Moraes2005}, a $-z_i/L$ stability correction accounts for buoyancy effects in the surface layer \citep{Wyngaard1974, Panofsky1977, Hojstrup1982}, and the variance is considered constant  relative to $w_\star$ in the mixed layer \citep{Caughey1979, Zilitinkevich2006, BouZeid2018}. Despite its integral role in characterising the three-dimensional (3D) structures of large-scale rolls and cells, the behaviour of $\langle v'^2 \rangle$ across the full range of convective conditions remains less investigated than its streamwise and vertical counterparts.

In this study, simulations across the parameter space of $-z_i/L$ are used to further investigate the variance formulations in tables~\ref{tab:vertical} and \ref{tab:streamwise}, and to evaluate the scaling of turbulent statistics with $u_*$ and $w_\star$. Particular attention is paid to the spanwise velocity, which has historically received less attention within sheared convection literature. The objective is to reconcile the statistical observations with the structural transition between thermal rolls and cells. This is achieved by assessing contributions to the variance budget equations, including scale-wise spectral budgets. The budget analysis reveals the pathways of energy as it is produced and distributed across heights and velocity components, highlighting the important role of pressure in this process. The remainder of the work is organised as follows: section~\ref{sec:methods} outlines the methodology and numerical framework. Sections~\ref{sec:variance} and \ref{sec:variancebudgets} present the structural evolution of the variance profiles and the accompanying budget analyses. Finally, section~\ref{sec:conclusions} summarises the results and findings.

\section{Methodology}
\label{sec:methods}

\subsection{Large-eddy simulations}

A suite of thirteen high-resolution large-eddy simulations (LES) is used to investigate, in a systematic and quantitative way, the structural and statistical evolution of turbulent variances as convective instability increases from a shear-dominated boundary layer to the absolute limit of penetrative free convection. The first six cases in weaker convection were previously published \citep{Heisel2024}, and this study expands the dataset to simulate the remainder of the parameter space. The numerical discretisation and boundary conditions remain consistent to ensure statistical fidelity.

The LES flow solver and numerical methods used in this study are described by \citet{Heisel2024} and are based on a modified version of the code developed by Albertson and co-authors \citep{Albertson1996, Albertson1999, PorteAgel2000}. The solver utilises a staggered vertical grid for the wall-normal velocity component to optimise vertical differentiation via second-order-accurate finite differencing. Horizontal derivatives are computed spectrally with full de-aliasing using the $3/2$ zero-padding rule. The flow is advanced in time using a second-order-accurate Adams--Bashforth integration scheme. Boundary conditions are periodic in the horizontal directions, with zero penetration enforced at the bottom of the computational domain; the upper boundary employs a stress-free condition. Fluctuations in the top 25\% of the domain are systematically relaxed using a Rayleigh damping layer to achieve the upper boundary conditions and mitigate gravity wave reflections \citep{Nieuwstadt1993}. 

The estimation of subgrid-scale (SGS) stresses uses a Lagrangian averaged scale-dependent (LASD) dynamic model to define the SGS eddy viscosity \citep{BouZeid2005}. The SGS heat flux is closed using the same dynamic formulation alongside a constant subgrid turbulent Prandtl number, $\Pran_{\mathrm{sgs}}=0.4$, with a comprehensive technical description of the numerical framework detailed elsewhere \citep{Kumar2006, Salesky2017}. The surface stress in this wall-modelled LES framework is estimated locally in space and time using MOST \citep{Moeng1984}. To mitigate log-law mismatch and wall-model overshoot, the wall model evaluates surface stresses using velocity values at $0.05z_i$ instead of the first grid point \citep{Heisel2024}.

The simulation parameters and resulting bulk characteristics are summarised in table~\ref{tab:cases}. Following \citet{Heisel2024}, all simulations employ an aerodynamic surface roughness length of $z_0 = 0.1\,\mathrm{m}$, an initial reference potential temperature of $\theta_0 = 300\,\mathrm{K}$, an imposed stably-stratified capping inversion with $0.003\,\mathrm{K\,m^{-1}}$ temperature gradient, and a constant Coriolis frequency of $f = 1\times10^{-4}\,\mathrm{s}^{-1}$. The latter two forcings are widely utilised to represent realistic atmospheric conditions \citep[see, e.g.,][]{Sullivan2011, Salesky2017, Salesky2018}. The imposed geostrophic wind velocity, $U_g$, decreases gradually from the maximum value $U_g = 15.0\ \mathrm{m\,s}^{-1}$ that characterises the initial cases dominated by shear to the limit of penetrative free convection ($U_g = 0.0\ \mathrm{m\,s}^{-1}$). Penetrative convection in this context indicates the presence of an overlying stable capping inversion layer that bounds the convective mixed layer but can be penetrated by buoyant plumes. The imposed surface heat flux, $Q$, increases from $0.10\ \mathrm{K\,m\,s^{-1}}$ to $0.24\ \mathrm{K\,m\,s^{-1}}$ across the initial shear-dominated simulations, and is subsequently held constant at $0.24\ \mathrm{K\,m\,s^{-1}}$ for the moderately to highly convective cases as $U_g$ decreases. This results in a range $-z_i/L$ covering more than four orders of magnitude from $-z_i/L = 2.6$ up to effectively infinity for the zero-mean-wind case. The finite reported value of $5.7 \times 10^4$ for the free convection case (c13) in table~\ref{tab:cases} results from a small residual mean stress with $u_* = 0.02\ \mathrm{m\,s^{-1}}$ arising from imperfect statistical convergence of the momentum flux.

In the cases with weaker convection (c1--c6), a large area of $12\times12\ \mathrm{km}$ is simulated in the horizontal and $2\ \mathrm{km}$ in the vertical. This horizontal domain corresponds to approximately $10z_i$ to $12z_i$, consistent with previous LES studies \citep{Salesky2017, Salesky2018, deRoode2022, Greene2025}, which is required to avoid any artificial constriction of the highly elongated turbulent roll structures that scale with $z_i$.

\begin{table}
    \centering
    \def~{\hphantom{0}}

    \begin{adjustbox}{max width=\textwidth}
    \begin{tabular}{cccccccc}
        Case &
        \makecell{$U_g$ ($\mathrm{m\,s^{-1}}$)} &
        \makecell{$Q$ ($\mathrm{K\,m\,s^{-1}}$)} &
        \makecell{Domain Size\\$L_x \times L_y \times L_z$ ($\mathrm{km}$)} &
        \makecell{Grid Size\\$N$ (points)} &
        \makecell{Turnover\\Times ($z_i/w_\star$)} &
        \makecell{Turnover\\Times ($z_i/u_*$)} &
        $-z_i/L$ \\
        \hline
        \multicolumn{8}{c}{\citet{Heisel2024}} \\
        \hline
        c1  & 15   & 0.10 & $12\times12\times2$ & $1024\times1024\times512$ & 5.2  & 2.8 & 2.6 \\
        c2  & 15   & 0.17 & $12\times12\times2$ & $1024\times1024\times512$ & 6.0  & 2.7 & 4.3 \\
        c3  & 15   & 0.24 & $12\times12\times2$ & $1024\times1024\times512$ & 6.7  & 2.7 & 6.0 \\
        c4  & 12   & 0.24 & $12\times12\times2$ & $1024\times1024\times512$ & 6.6  & 2.3 & 9.6 \\
        c5  & 9    & 0.24 & $12\times12\times2$ & $1024\times1024\times512$ & 8.8  & 2.5 & 18 \\
        c6  & 6    & 0.24 & $12\times12\times2$ & $1024\times1024\times512$ & 6.6  & 1.4 & 40 \\
        \hline
        \multicolumn{8}{c}{Present Study} \\
        \hline
        c7  & 5    & 0.24 & $5.76\times5.76\times2$ & $480\times480\times500$ & 20 & 3.9 & 55 \\
        c8  & 4    & 0.24 & $5.76\times5.76\times2$ & $480\times480\times500$ & 23 & 3.9 & 84 \\
        c9  & 3    & 0.24 & $5.76\times5.76\times2$ & $480\times480\times500$ & 27 & 3.8 & 140 \\
        c10 & 2    & 0.24 & $5.76\times5.76\times2$ & $480\times480\times500$ & 31 & 3.5 & 260 \\
        c11 & 1    & 0.24 & $5.76\times5.76\times2$ & $480\times480\times500$ & 34 & 2.7 & 810 \\
        c12 & 0.5  & 0.24 & $5.76\times5.76\times2$ & $480\times480\times500$ & 38 & 2.1 & 2400 \\
        c13 & 0    & 0.24 & $5.76\times5.76\times2$ & $480\times480\times500$ & 38 & 0.7 & 57\,000 \\
        \hline
    \end{tabular}
    \end{adjustbox}
    \caption{Summary of the six previous and seven new LES cases, including the fixed geostrophic wind speed $U_g$, fixed surface heat flux $Q$, duration relative to both convective ($z_i/w_\star$) and shear ($z_i/u_*$) time scales, and the resulting bulk stability parameter $-z_i/L$.}
    \label{tab:cases}
\end{table}

Following the structural shift from rolls to cellular convection, the large-scale structures are less elongated in the horizontal and an $O(10z_i)$ domain length is no longer required. Accordingly, the grid dimensions were adjusted for the strongly convective regime (c7--c13) to maintain computational efficiency. While the horizontal physical domain was reduced from $12~\mathrm{km}$ to $5.76~\mathrm{km}$ (${\sim}5z_i$), the spatial resolution was preserved by adjusting the grid-point allocation from $1024 \times 1024 \times 512$ to $480 \times 480 \times 500$. The minor adjustment in vertical grid points was required for parallel partitioning of the new simulations on the utilised computing infrastructure. The vertical grid spacing for all cases is $\rmDelta z/z_i \lesssim 0.004$. Case c6 was repeated using both domains to ensure the results are not affected by the smaller domain size. Appendix~\ref{sec:appA} shows that variance statistics closely match between the two domains.

All cases used an iterative spin-up procedure to mitigate inertial oscillations during the simulation \citep{Heisel2024}. The post-spin-up duration of each simulation varied from 60 physical minutes for case c1 to 5.5 hours for case c13. The goal was to capture sufficient turbulent events related to both buoyancy (whose relevant timescale is $z_i/w_\star$) and shear ($z_i/u_*$ timescale). This was assessed by the post-spin-up turnover times in table~\ref{tab:cases} and the statistical convergence of the heat and momentum flux profiles. The long durations under strong convection were required to sufficiently capture the relatively infrequent, prolonged shear-related events and accurately evaluate their diminishing strength. 

\begin{figure}
  \centering
  \includegraphics[width=1.0\textwidth]{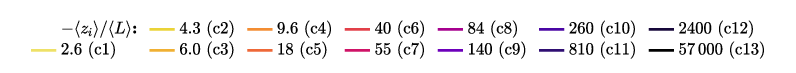} 
  \vspace{0.2cm}
  \includegraphics[width=1.0\textwidth]{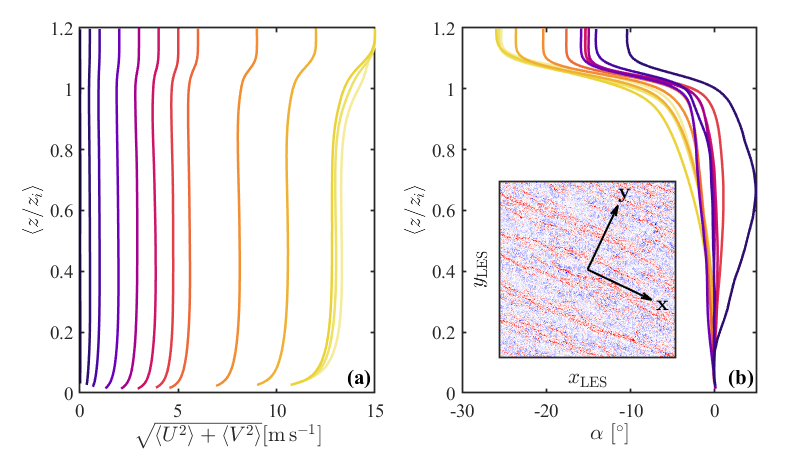}
  \caption{Vertical profiles of the mean wind properties for each LES case in table~\ref{tab:cases} plotted as a function of normalised height $\langle z/z_i \rangle$: (\textit{a}) magnitude; (\textit{b}) direction $\alpha$ relative to the LES grid. In all subsequent results, the $x{-}y$ coordinate system is aligned with the near-surface wind direction, as shown in the inset of (\textit{b}), which shows instantaneous $w'$ in a horizontal plane at $z/z_i=0.1$ for the weakest convective case c1.}
  \label{fig:1}
\end{figure}

Figure~\ref{fig:1}(\textit{a}) shows the mean wind magnitude for each LES case. In this and all later results, statistical moments are averaged across horizontal position and time. The profiles indicate the presence of a well-developed mixed layer, which is a common feature of the CBL \citep{Stull1988}. Within this layer, buoyancy-driven thermal plumes efficiently redistribute momentum and potential temperature, producing nearly uniform mean profiles. Above this, there is an entrainment layer between the CBL and the capping inversion where the wind magnitude reaches $U_g$.

In a rotating atmosphere governed by Coriolis forces, the mean wind vector typically changes direction with height through the planetary boundary layer \citep{Tennekes1972,Grossman1982}. Throughout most of the CBL, this veering is mitigated by vertical mixing, as seen in the figure~\ref{fig:1}(\textit{b}) profiles of wind direction $\alpha$. To account for the mismatch between the wind direction within the CBL and the imposed $U_g$ aligned with the LES grid, the horizontal velocity components and their statistics are rotated according to the surface wind direction in this study. Because the wind direction is relatively constant throughout the surface and mixed layers (i.e. $\pm 5^\circ$ for $z/z_i \lesssim 0.85$), a single fixed rotation by the mean surface direction at $z/z_i \leq 0.05$ is applied across the full vertical domain. This fixed rotation provides a consistent streamwise--spanwise ($x{-}y$) coordinate system that preserves the geometric orientation of boundary-layer-spanning structures across height while providing a physically meaningful decomposition of the turbulent fluctuations aligned with the near-surface mean flow and shear \citep{Wilczak2001}. The rotation aligns the coordinate system with the orientation of the dominant turbulent structures, as illustrated by the elongated rolls in the figure~\ref{fig:1}(\textit{b}) inset.

\begin{figure}
  \centering
  \includegraphics[width=1.0\textwidth]{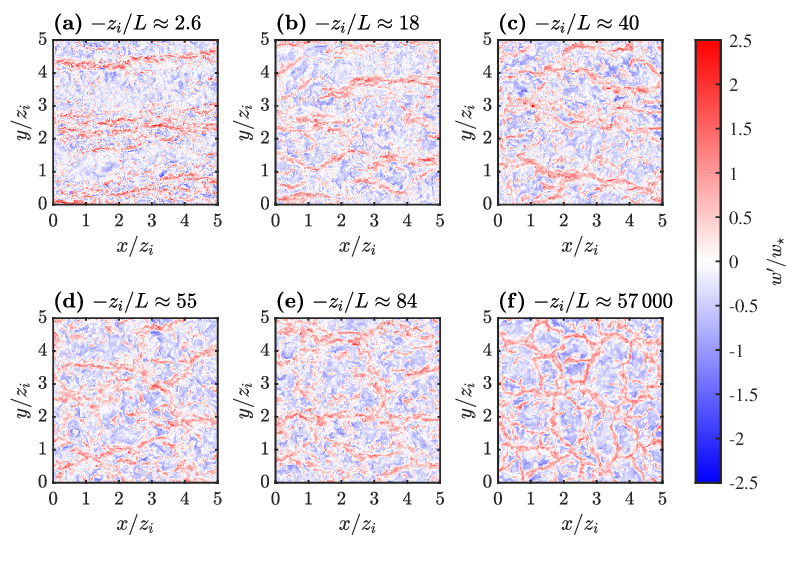}
  \caption{Instantaneous fluctuations of normalised vertical velocity $w'/w_\star$ in the horizontal plane at $z/z_i=0.1$ across six cases to visualise the transition from thermal rolls to cellular plumes: (\textit{a}) c1, (\textit{b}) c5, (\textit{c}) c6, (\textit{d}) c7, (\textit{e}) c8, (\textit{f}) c13. In each case the plane is rotated to align with the surface wind direction and confined to the same area relative to $z_i$.}
  \label{fig:2}
\end{figure}

The inset fluctuating field is shown for an additional five cases in figure~\ref{fig:2} using the rotated coordinate system. The instantaneous horizontal cross-sections at $z=0.1z_i$ illustrate the change in coherent turbulence structure with increasing instability: the weakly convective near-surface flow is populated by longitudinal convective rolls that organise the flow into stable, counter-rotating circulations \citep{LeMone1976}. As instability increases, these large-scale structures gradually break down into a mixed roll–-cellular organisation comprising predominantly buoyancy-driven cellular plume structures. Previous studies report that cells become dominant around $-z_i/L \approx 20\text{--}25$ \citep{Grossman1982,Weckwerth1999,Salesky2017} based on the visual end of roll and cell coexistence \citep{Grossman1982}, the loss of linear organisation \citep{Weckwerth1999}, and a geometric roll factor \citep{Salesky2017}. Here, the absence of longitudinal coherence associated with rolls occurs for $-z_i/L\approx 50$ between figures~\ref{fig:2}(\textit{c}) and \ref{fig:2}(\textit{d}) based on visual observation and the same roll factor. This higher threshold is further supported by later quantitative results. The later transition relative to $-z_i/L$ may be due to a number of methodological factors: the impact of inertial oscillations and wall stress overshoot on $u_*$ and $L$ \citep{Heisel2024}, which have both been mitigated here; the substantial difference in spatial resolution; and different grid aspect ratios. Considering the uncertainty from these factors, the present value $-z_i/L\approx50$ is used only to differentiate trends before and after the transition, and no conclusions are made regarding a universal $-z_i/L$ value for the transition.

\subsection{Temporal averaging and normalisation}
\label{sec:normalisation}

For convective conditions with a stable capping inversion, the CBL slowly grows as the inversion is entrained into the boundary layer. This growth is significant over the duration of the highly convective cases. For example, the capping inversion depth, $z_i$, grows by up to $65\%$ in c12. In the setup of each case, the initial capping inversion was positioned to avoid growth of the CBL into the damping layer by the end of the simulation. An important additional consideration is that the key scaling parameters, $z_i$ and $w_\star$, evolve as the simulation progresses. 

\begin{figure}
  \centering
  \includegraphics[width=1.0\textwidth]{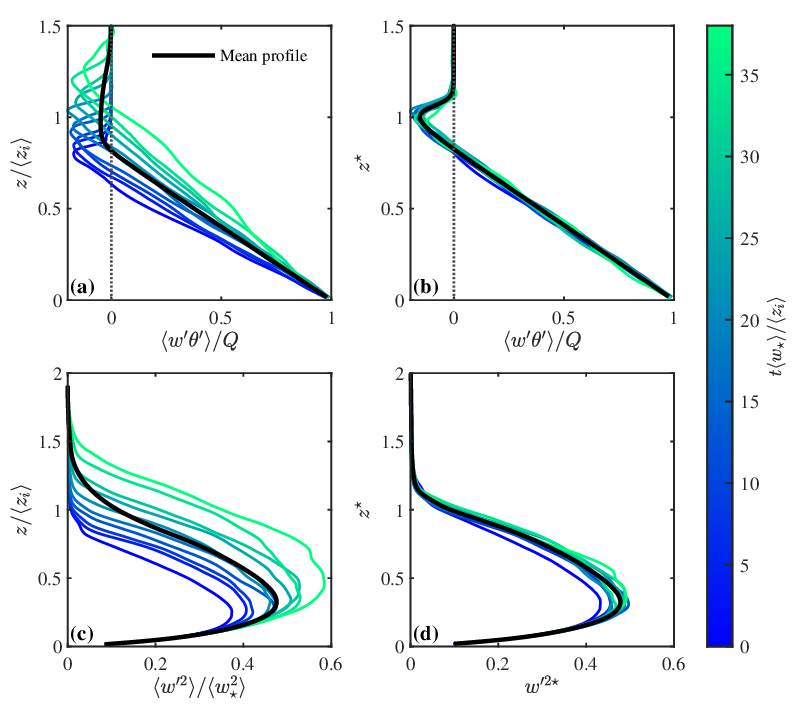}
  \caption{Profiles of total heat flux $\langle w'\theta' \rangle$ (top row) and vertical velocity variance $\langle w'^2 \rangle$ (bottom row) at different time intervals during case c12, including the time-averaged profile (black curve). In (\textit{a,c}), statistics are normalised by the overall time-averaged CBL depth $\langle z_i\rangle$ and convective velocity $\langle w_\star\rangle$. In (\textit{b,d}), local-in-time scaling is applied before temporal averaging using the instantaneous parameters $z_i(t)$ and $w_\star(t)$. In this and later figures, the star (${}^\star$) superscript denotes local-in-time normalised statistics, e.g. $z^\star=\langle z/z_i(t) \rangle$ and $w'^{2\star}=\langle w'^2/w_\star^2(t) \rangle$.}
  \label{fig:3}
\end{figure}

Figure~\ref{fig:3} demonstrates the influence of the growing CBL on time-averaged statistics using case c12. The coloured lines represent independent profiles averaged within ten one-minute intervals spaced throughout the simulation time $t$, while the solid black curve denotes the overall time-averaged profile. The growth over time of the CBL is evident in figure~\ref{fig:3}(\textit{a,c}), noting the capping inversion depth $z_i$ is defined as the height of the minimum in the kinematic heat flux profile. This growth leads to a smoothing of the time-averaged profile, particularly near the inversion. In contrast, applying local-in-time scaling using $z_i(t)$ and $w_\star(t)$ produces substantially improved profile collapse in figure~\ref{fig:3}(\textit{b,d}), with the individual time intervals remaining closely aligned with the overall mean. The resulting collapse indicates that local-in-time normalisation accounts for the evolving convective scales and avoids contamination of statistics due to non-stationary CBL growth. Accordingly, all statistics in this study -- including for figures~\ref{fig:1} and \ref{fig:2} -- are normalised using local-in-time parameters before averaging across time. Here, the star (${}^\star$) superscript denotes normalisation using the instantaneous convective scales. The normalised vertical coordinate is defined as $z^\star = \langle z/z_i(t) \rangle$, and the normalised velocity variances are $w'^{2 \star} = \langle w'^2/w_\star^2(t) \rangle$, where time-averaging indicated by brackets occurs after normalisation. Owing to the CBL growth, the $z^\star$ points represented by the LES grid change over time. For computing statistics as a function of $z^\star$, the local-in-time statistics were first interpolated onto a common $z/z_i$ grid before averaging the statistics across time.

\subsection{Subgrid-scale variance statistics}
\label{sec:sgs}

The SGS model of the LES solver does not directly estimate the SGS variance (i.e. normal stresses), and the corresponding SGS normal stresses are instead grouped with the pressure term \citep{Albertson1996}. Because the SGS energy can be substantial near the surface where characteristic turbulent length scales are restricted, the SGS contribution to the variance statistics is estimated in post-processing for all cases to ensure the results are independent of the grid resolution.

The SGS variance estimation extrapolates the energy to subgrid scales using the Leonard stress and assuming a $-5/3$ inertial subrange \citep{Knaepen2002,Salesky2012}.  The SGS variances for $u$ and $v$ are transformed using the same coordinate rotation described above for the resolved statistics. The SGS contributions are then added to the resolved quantities prior to normalisation and time averaging.

The SGS estimation relies on the grid resolution being within the inertial subrange of scales exhibiting a $-5/3$ power law in the spectrum. To evaluate this criterion, SGS variances were estimated using a range of grid and test filter sizes to compute the Leonard stress, where consistency across filters indicates the assumption is valid and the estimate is reliable. Variance statistics are only presented for points above a cutoff height where the filter comparisons converge, and near-surface points with discrepancies across filter sizes are excluded. Further details of the SGS estimate and the determination of the cutoff height are provided in Appendix~\ref{sec:appB}.

\subsection{Resolved pressure statistics}
\label{sec:pressure}

As noted above, kinematic pressure statistics from the LES solver include the SGS kinetic energy. In comparison with literature \citep{Ding2018}, it was found that some statistics derived from this pressure such as the variance compare well with the true pressure, while statistics dependent on smaller scales such as the pressure--strain correlation show substantial discrepancies. Correcting the pressure using estimates of the instantaneous SGS energy is not reliable because the SGS estimates only work well for the average contribution \citep{Knaepen2002}.

Instead, the resolved pressure field is determined offline via the Poisson equation derived from the Navier--Stokes equations \citep{Moeng1986,Yi2025}, excluding the SGS stress term. Previous studies demonstrated that the SGS stress contribution to the resolved pressure field is small to negligible for these high-Reynolds-number turbulent statistics \citep{Andren1994,Mironov2000,Heinze2016}. The boundary conditions for the pressure solver are zero gradient at the surface and zero pressure at the upper boundary \citep{Heinze2016}. For each case, the resolved pressure was computed from 60 instantaneous flow volumes evenly spaced across the simulation duration. The resolved pressure trends align with previous literature, and corresponding pressure statistics closely match with indirect estimates from the residual of the variance budgets described below.

\subsection{Variance budget equations}
\label{sec:budgets}

The subsequent analysis evaluates the variance budget equations to quantify the production, transport, and redistribution of turbulent energy across the CBL under varying convection. These equations and their notation are introduced here to facilitate this assessment. Upon Reynolds decomposition and averaging of the Navier--Stokes equations, the variance budget equations are written as \citep{Stull1988, Garratt1992}:
\begin{equation}\label{eq:3}
\frac{\partial \langle u'^2 \rangle}{\partial t} = \underbrace{-2\langle u'w' \rangle\frac{\partial U }{\partial z}}_{\mathcal{P}_u} - \underbrace{\frac{\partial \langle u'^2 w' \rangle}{\partial z}}_{\mathcal{T}_u} + 2f \langle u' v' \rangle + \underbrace{2 \langle p' \frac{\partial u'}{\partial x} \rangle}_{\mathcal{R}_u} - \varepsilon_{u},
\end{equation}
\begin{equation}\label{eq:4} 
\frac{\partial \langle v'^2 \rangle}{\partial t} = \underbrace{-2\langle v'w' \rangle\frac{\partial V}{\partial z}}_{\mathcal{P}_v} - \underbrace{\frac{\partial \langle v'^2 w' \rangle}{\partial z}}_{\mathcal{T}_v} - 2f \langle u' v' \rangle + \underbrace{2 \langle p' \frac{\partial v'}{\partial y} \rangle}_{\mathcal{R}_v} - \varepsilon_{v},
\end{equation}
\begin{equation}\label{eq:5} 
\frac{\partial \langle w'^2 \rangle}{\partial t} = \underbrace{2\frac{g}{\theta_0}\langle w'\theta' \rangle}_{\mathcal{B}_w} - \underbrace{\frac{\partial \langle w'^3 \rangle}{\partial z}}_{\mathcal{T}_w} - \underbrace{2\frac{\partial \langle w'p' \rangle}{\partial z}}_{\Pi_w} + \underbrace{2\langle p'\frac{\partial w'}{\partial z} \rangle}_{\mathcal{R}_w} - \varepsilon_{w}.
\end{equation}
Here, $(\cdot)'$ denotes the fluctuating component of a given variable, $U=\langle u\rangle$ and $V=\langle v\rangle$ denote the mean streamwise and spanwise velocities, respectively, and $p$ is the resolved pressure computed offline as described above. The vertical velocity variance incorporates buoyancy effects through the Boussinesq approximation, consistent with the LES solver.

The factors analysed herein are the production of energy by shear ($\mathcal{P}_u$ and $\mathcal{P}_v$), buoyancy ($\mathcal{B}_w$), vertical transport by turbulence ($\mathcal{T}_i$) and pressure ($\Pi_w$), pressure--strain correlation or redistribution by pressure ($\mathcal{R}_i$), and SGS dissipation of turbulent energy ($\varepsilon_i$). In the LES solver, the total TKE dissipation rate $\varepsilon = \varepsilon_u + \varepsilon_v + \varepsilon_w = -\langle \tau_{ij}^{\mathrm{sgs}} \tilde{S}_{ij} \rangle$ is determined directly from the LASD Smagorinsky model, where $\tau_{ij}^{\mathrm{sgs}}$ and $\tilde{S}_{ij}$ are the deviatoric SGS stress tensor and resolved strain-rate tensor respectively. In later results, dissipation is presented as the total $\varepsilon$.

Besides the dissipation, each term is computed from the resolved LES variables with the exceptions noted here. The computed production terms are based on the total fluxes including the SGS contribution. Similarly, the transport $\mathcal{T}$ includes the correlation between the resolved velocity and SGS stresses \citep{Andren1994, Mironov2000}. The redistribution $\mathcal{R}$ acts to transfer energy between velocity components and for incompressible flow has a zero-trace pressure--strain tensor ($\mathcal{R}_u + \mathcal{R}_v + \mathcal{R}_w = 0$), such that any energy removed from one variance is injected into the other two \citep{Pope2000}. The storage term on the left-hand side of (\ref{eq:3})--(\ref{eq:5}) is small but non-zero in the present simulations owing to the slow growth of the CBL. The redistribution of energy between $u'$ and $v'$ by Coriolis ($\pm 2f \langle u' v' \rangle$) is also negligible relative to the other terms and is not shown in the later results.

\section{Variance statistics}
\label{sec:variance}

\subsection{Convective similarity}
\label{sec:convectivesimilarity}

\begin{figure}
  \centering
  \includegraphics[width=1.0\textwidth]{figures/global_legend.eps} 
  \vspace{0.2cm}
  \includegraphics[width=1.0\textwidth]{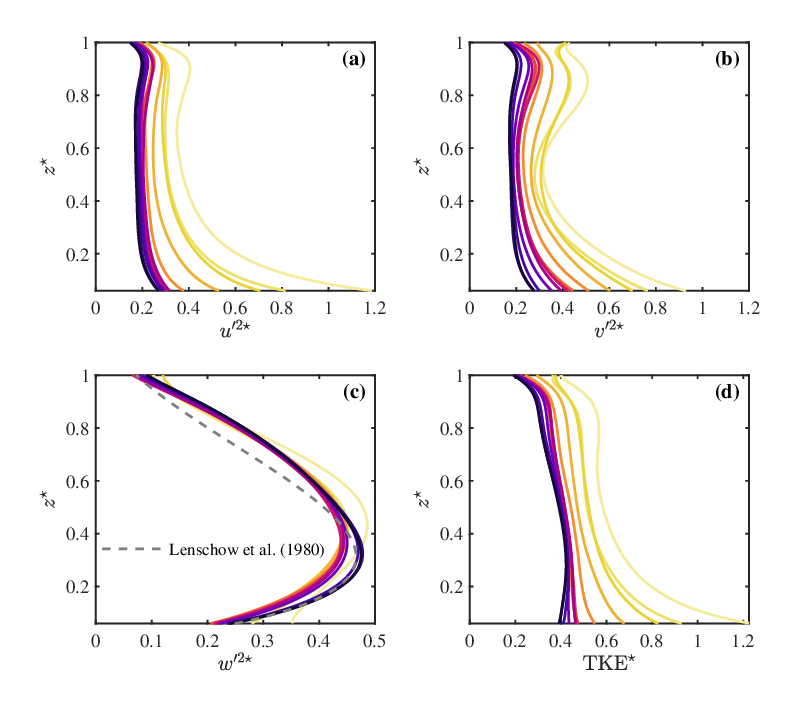}
  \caption{Profiles of the total (resolved + SGS) turbulent energy: (\textit{a}) streamwise variance $u'^{2\star}$, (\textit{b}) spanwise variance $v'^{2\star}$, (\textit{c}) vertical velocity variance $w'^{2\star}$, and (\textit{d}) turbulent kinetic energy ($\mathrm{TKE}^\star$). The dashed grey curve in (\textit{c}) shows the empirical parametrisation of \citet{Lenschow1980} from table~\ref{tab:vertical}.}
  \label{fig:4}
\end{figure}

The variance statistics are first evaluated relative to the convective velocity $w_\star$ to identify overall trends including adherence to traditional free convection scaling. Figure~\ref{fig:4} presents profiles of the total velocity variances and TKE across the thirteen LES cases, where each profile includes the resolved and estimated SGS component as described in section~\ref{sec:sgs}. A comparison of figure~\ref{fig:4}(\textit{a}) and figure~\ref{fig:4}(\textit{b}) reveals that $u'$ is the dominant velocity component near the surface under weak convection for case c1. However, $v'$ consistently exceeds $u'$ throughout most of the CBL under moderately to strongly convective conditions, with $v'>u'$ most evident for cases c4--c10 spanning the transition from rolls to cells. The spanwise variance also increases more than $u'$ in the upper half of the CBL, corresponding to the top of the counter-rotating thermal rolls. As buoyancy becomes increasingly dominant, both horizontal velocity variances decrease across all heights to the same value of $0.18$ for case c13, which is approximately half the $B_{u_3}=0.35$ value in table~\ref{tab:streamwise}. Both variances continuously decrease without collapsing onto a common curve, indicating that a universal profile with mixed-layer scaling $w_\star$ is only approached asymptotically for the horizontal velocities under strong convection. 

In contrast, the normalised vertical velocity in figure~\ref{fig:4}(\textit{c}) exhibits limited variability across the parameter space and is in generally good agreement with the empirical mixed-layer parametrisation proposed by \citet{Lenschow1980} throughout much of the mixed layer, particularly approaching free convection. The largest deviations occur in the upper half of the CBL where capping inversion and entrainment effects may be relevant. Despite the better collapse, there is a systematic trend in which the variance peak initially decreases under weakly convective conditions before increasing with instability, accompanied by a downward shift in its vertical location. The initial decrease in the $w'^{2\star}$ peak occurs while shear-generated turbulence is still dynamically important, whereas the subsequent increase begins only after shear production has become relatively weak. A similar qualitative trend has been observed in previous simulations \citep{Zhou2019}, although its dynamical origin has not been explicitly examined.

The corresponding normalised total turbulent kinetic energy $\mathrm{TKE}^\star=\frac{1}{2}(u'^{2\star}+v'^{2\star}+w'^{2\star})$ shown in figure~\ref{fig:4}(\textit{d}) provides a useful measure of the combined response of the three velocity components to increasing convection. Following an initial decrease under weak convection, $\mathrm{TKE}^\star$ becomes approximately invariant near the bottom of the mixed layer, around $z^\star\approx0.4$, beginning with cases c6--c7. This behaviour coincides with the transition to predominantly cellular structures at $-z_i/L\approx 50$ discussed previously. It is also the point at which the mixed-layer TKE first becomes comparable to, and locally exceeds, the near-surface value. In these conditions, shear production at the surface has already been substantially reduced and buoyant plumes within the mixed layer become the dominant energy-containing motions. The energy $\mathrm{TKE}^\star$ continues to decrease in the lower and upper limits of the CBL until free convection is reached. This decrease is primarily associated with the weakening of the horizontal velocity fluctuations in figure~\ref{fig:4}(\textit{a,b}) as the roll and cell structures progressively lose their horizontal circulations at the top and bottom of the CBL. The invariance in the mixed layer indicates that the reduction in $u'$ and $v'$ are offset by the increased $w'$ at these heights.

Taken together, the results in figure~\ref{fig:4} indicate that mixed-layer scaling with $w_\star$ captures the leading-order magnitude of the variances and TKE under strongly convective conditions, but does not completely describe the evolution of the profiles with changing convection. The remaining variability is attributed to the changing organisation of the largest turbulent structures and its impact on the distribution of turbulent energy, motivating the later examination of variance budget contributions.

\begin{figure}
  \centering
  \includegraphics[width=0.75\textwidth]{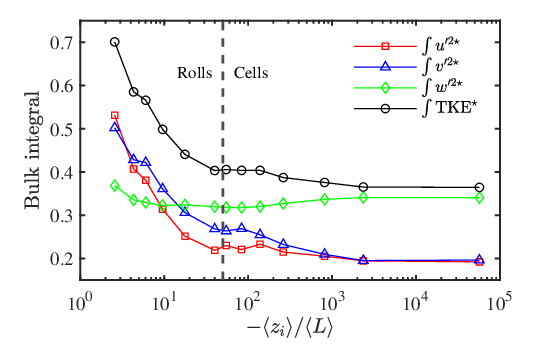}
  \caption{Variance and TKE statistics integrated across the CBL depth and shown as a function of global instability $-z_i/L$. The dashed line at $-z_i/L=50$ indicates the approximate point where cells become dominant over rolls.}
  \label{fig:5}
\end{figure}

The non-monotonic $w'^{2\star}$ peak and higher spanwise variance $v'>u'$ in figure~\ref{fig:4} can be further assessed by summing the variances across the full CBL depth using integration. Depth integration removes the detailed vertical structure while retaining the total contribution of each velocity component to the boundary-layer turbulent energy. The integral is evaluated from the lowest valid grid level above the surface up to the top of the domain to include any residual turbulence in the entrainment layer above $z_i$. Figure~\ref{fig:5} shows the depth-integrated variances and TKE, where the integral notation in the legend is shorthand for $\int_0^\infty u'^{2\star}\,\mathrm{d}z^{\star}$. 

The integrated vertical variance $\int w'^{2\star}$ exhibits the same trends as the peak in figure~\ref{fig:4}: while the value varies weakly across four orders of magnitude in $-z_i/L$, there is an initial decrease followed by a gradual increase approaching the free-convection limit. Figure~\ref{fig:5} reveals that the minimum coincides with the end of the transition from rolls to cells (dashed line), indicating a connection between the change in large-scale structure and the evolving magnitude of $w'^{2\star}$ which is further explored in section~\ref{sec:variancebudgets}.

The comparison of horizontal velocities in figure~\ref{fig:4} is more apparent from the integrated statistics in figure~\ref{fig:5}, where $\int v'^{2\star}$ exceeds $\int u'^{2\star}$ from approximately case c2 onward. The difference decreases progressively until a common value is reached under near-free convection conditions. Further, $v'$ is the most energetic velocity component across the CBL for cases c2--c4, despite the absence of turbulence production along $v'$.

The integrated TKE decreases substantially across the longitudinal roll regime as shear-generated turbulence diminishes. The TKE then decreases weakly in the cell regime, corresponding to the reduction of $u'$ and $v'$ at the CBL top and bottom observed in figure~\ref{fig:4}. These depth-integrated results therefore reinforce the observations from the figure~\ref{fig:4} profiles: the transition to cellular convection changes how turbulent energy is distributed among the velocity components. The nature of these changes is evaluated further in section~\ref{sec:variancebudgets} using the variance budgets.

\subsection{Surface layer similarity}
\label{sec:surfacesimilarity}

\begin{figure}
  \centering
  \includegraphics[width=1.0\textwidth]{figures/global_legend.eps}
  \vspace{0.2cm}
  \includegraphics[width=1.0\textwidth]{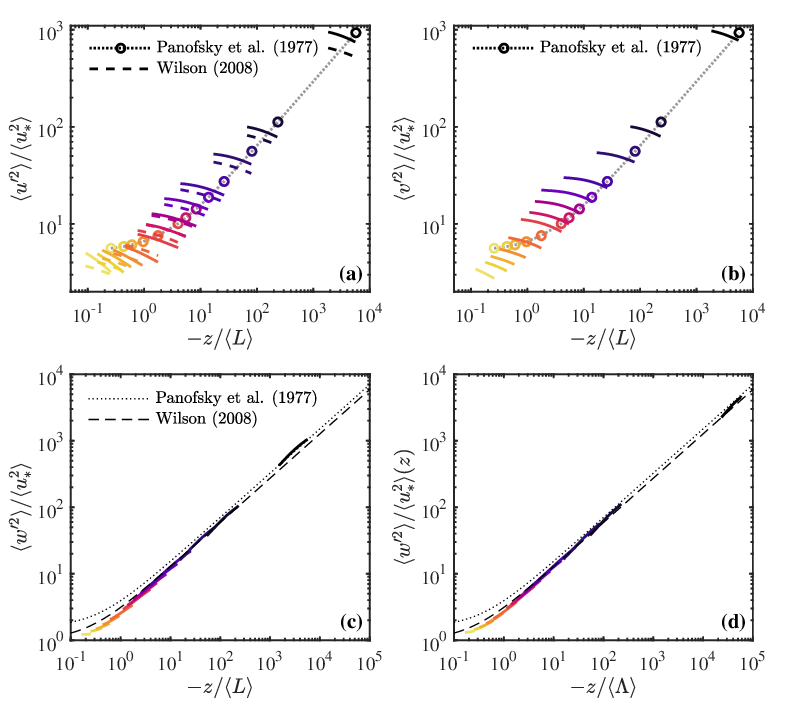}
  \caption{Comparison of surface layer variances with Monin--Obukhov similarity: (\textit{a,b,c}) streamwise, spanwise, and vertical velocities, respectively, normalised by $u_*$ and shown as a function of $-z/\langle L\rangle$ based on the surface Obukhov length in \eqref{eq:1}; (\textit{d}) same as (\textit{c}), but with $u_*(z)$ and the local Obukhov length $\Lambda$ based on local-in-height momentum and heat fluxes. Empirical MOST formulations from tables~\ref{tab:vertical} and \ref{tab:streamwise} are included for comparison \citep{Panofsky1977,Wilson2008}. The \citet{Panofsky1977} points in (\textit{a,b}) are for 0.1$z_i$, which is the highest position included for each profile.}
  \label{fig:6}
\end{figure}

Section~\ref{sec:convectivesimilarity} presented statistics across the CBL with respect to the convective scales. Here, the focus is on the transition between $u_*$ and $w_\star$ scaling within the surface (log) layer. The surface-layer variances are first examined using MOST in figure~\ref{fig:6}, where the competing effects of surface shear and buoyancy are expressed through $u_*$ and the Monin--Obukhov stability parameter $\zeta$. The height range shown for each case extends from the SGS cutoff height defined in Appendix~\ref{sec:appB} to an upper bound of $0.1z_i$.

From case to case, the streamwise and spanwise velocity variances in figures~\ref{fig:6}(\textit{a}) and \ref{fig:6}(\textit{b}) increase with convection when normalised by $u_*$. Both velocities align with a 2/3 power law across cases under strong convection, consistent with the $w_\star^2 \sim u_*^2 (-z_i/L)^{2/3}$ scaling. The $v'$ profiles in figure~\ref{fig:6}(\textit{b}) exhibit this power law even for weak convection, indicating that the near-surface spanwise variance in this regime is dominated by convective-scaled roll motions that span the CBL \citep{Khanna1997}. The \citet{Panofsky1977} relation aligns well with the $w_\star$-scaled regime in strong convection, but does not accurately capture the transition from $u_*$ in weak convection.

As expected, similarity in figure~\ref{fig:6}(\textit{a,b}) using MOST is prevented by the height dependence within each profile. The \citet{Wilson2008} empirical fit captures the height dependence reasonably well for cases c7--c13 in the presence of cells, but underestimates the height dependence for cases c1--c6 with rolls. Thus, despite qualitatively capturing the trends for strong convection, an alternative formulation to MOST and the empirical fits is required to account for horizontal variance statistics in weak convection dominated by longitudinal rolls.

In contrast, the vertical velocity variance in figure~\ref{fig:6}(\textit{c}) yields a reasonable collapse of the profiles across the parameter space, indicating that MOST captures the transition in $\langle w'^2\rangle$ from shear to buoyancy-dominated scaling. The profiles are also qualitatively consistent with the empirical fits reported in the literature for the convective surface layer and mixed layer, including the approach to free convection \citep{Panofsky1977,Wilson2008,Wyngaard1971}.

Although the adherence of the vertical velocity variance to MOST is well established, uncertainty remains in the precise formulation and fitted constants, particularly under weak convection. For example, \citet{Hogstrom1990} found that the formulation of \citet{Panofsky1977} overestimated normalised vertical velocity fluctuations by approximately 25\% relative to reanalysed field measurements. Additionally, both previous LES and the present profiles exhibit lower variance values than the literature fits under weak and moderate convection \citep{Maronga2017}, with the weak-convection values falling below $B_w \approx 1.45\text{--}1.9$ predicted for neutral boundary layers (see, e.g., table~\ref{tab:vertical}). Owing to uncertainty in the total (resolved $+$ SGS) variances, no physical conclusions are drawn here regarding these lower values.

The collapse of $\langle w'^2\rangle$ improves further when local scaling is used \citep{Nieuwstadt1984}, as shown in figure~\ref{fig:6}(\textit{d}). The local scaling replaces the surface fluxes with local-in-height fluxes through the local friction velocity, $u_*(z)=[\langle u'w'\rangle(z)^2+\langle v'w'\rangle(z)^2]^{1/4}$, and the local Obukhov length, $\Lambda(z)=-u_*(z)^3\theta_0/[\kappa g\langle w'\theta'\rangle(z)]$, noting the magnitude of the $\langle v'w'\rangle$ contribution to $u_*(z)$ remains small ($\leq4\%$) compared to $\langle u'w'\rangle$ in the rotated coordinate system. The improved collapse obtained using $-z/\langle \Lambda \rangle$ indicates that the height-dependent local momentum and heat fluxes provide a more complete similarity of $\langle w'^2\rangle$, consistent with findings for neutral boundary layers \citep{Heisel2026}.

The results in figure~\ref{fig:6} therefore show that MOST captures the qualitative scaling trends across varying $z_i/L$ for all three velocity variance components, but complete similarity is precluded by underlying height dependencies. The collapse of profiles in figure~\ref{fig:6}(\textit{d}) is due to $w'$ being height-invariant in a neutral surface layer, such that the near-surface height dependence in figure~\ref{fig:4}(\textit{c}) is solely due to buoyancy effects that can be parameterised by $-z/L$. In the next section, the logarithmic profile in \eqref{eq:2} is used to separate the convection trends from the height dependence observed for the streamwise variance in figure~\ref{fig:6}(\textit{a}).

\subsection{Horizontal variances}
\label{sec:streamwisevariance}

\begin{figure}
  \centering
  \includegraphics[width=1.0\textwidth]{figures/global_legend.eps} 
  \vspace{0.2cm}
  \includegraphics[width=1.0\textwidth]{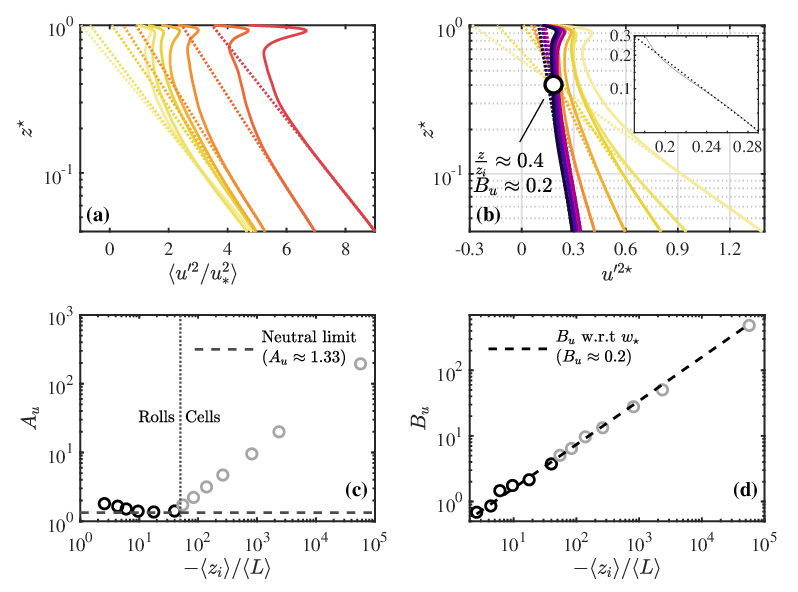}
  \caption{The streamwise velocity variance for weak convection fitted to $\langle u'^2 / u_*^2 \rangle = B_u - A_u \log{(z/0.4z_i)}$ within the surface layer. (\textit{a,b}) Profiles relative to $u_*$ and $w_\star$, respectively, where the dotted lines represent surface layer fits. The inset in (\textit{b}) evaluates the surface layer fit for the free-convection limit (case c13) to demonstrate the deviation from a strictly logarithmic profile. (\textit{c,d}) The fitted logarithmic slope $A_u$ and intercept $B_u$, respectively, as functions of global instability. The grey markers in (\textit{c,d}) indicate the cases deviating from the logarithmic variance profile.}
  \label{fig:7}
\end{figure}

\citet{Banerjee2015} predicted a logarithmic decay for the streamwise variance in weak convection, consistent with the attached eddy framework. Figure~\ref{fig:7}(\textit{a}) shows that the LES $u'^2$ profiles follow this logarithmic shape for cases c1--c6 with longitudinal rolls. Fits to \eqref{eq:2} (dotted lines) are evaluated over the same case-dependent height range as figure~\ref{fig:6}, extending from the SGS cutoff height defined in Appendix~\ref{sec:appB} to an upper bound of $0.1z_i$. A distinct logarithmic profile is observed within the surface layer for these cases, with limited change in the slope across cases but varying offset $B_u$ corresponding to a shift along the abscissa. The $u'$ profiles diverge from a strictly logarithmic decay beginning with case c7 and thereafter more closely resemble the shape of the $v'$ profiles and the \citet{Wilson2008} height dependence. The inset in figure~\ref{fig:7}(\textit{b}) evaluates the surface layer fit for the free-convection limit (case c13), visually demonstrating the departure from the logarithmic formulation under stronger convection.

The logarithmic fits use $0.4z_i$ as the reference height rather than $z_i$. This choice reflects the characteristic depth over which shear turbulence remains significant, noting the mixed layer is characterised by zero mean shear and constant variance (figure~\ref{fig:4}). The bottom of the mixed layer occurs near $0.4z_i$ \citep{Heisel2024}, although the height at which the mean shear approaches zero varies between approximately $0.3z_i$ and $0.55z_i$ among the cases. Using a representative value of $0.4z_i$ therefore provides a common reference for the shear-dominated region while preserving a simple two-parameter logarithmic formulation. Figure~\ref{fig:7}(\textit{b}) shows that the fitted profiles converge towards a common variance level near this height when expressed using $w_\star$, providing additional support for its interpretation as the relevant reference level for shear-generated turbulence in the CBL.

The fitted values of $A_u$ in figure~\ref{fig:7}(\textit{c}) indicate that in the presence of rolls, $A_u$ remains approximately constant to first-order accuracy. The constant value indicates that mechanical shear and wall-attached structures remain the dominant organising process for the $u'(z)$ height dependence. Once the stability parameter exceeds $-z_i/L\approx50$, the slope departs sharply from this first-order approximation to $A_u\sim (-z_i/L)^{2/3}$, coinciding with a shift from $u_*$ to $w_\star$ in governing the shape of $u'(z)$.

The fitted slopes $A_u \approx 1.36\text{--}1.8$ in the roll regime are somewhat larger than values reported for canonical neutral boundary layers \citep{Marusic2013,Banerjee2013,Qin2025}. One possible explanation is the finite horizontal extent of the present LES domain, which limits the longest resolved turbulent wavelengths. Such a restriction preferentially removes very-large-scale variance, with the effect becoming more important at increasing height as the characteristic eddy size grows, and may consequently produce a steeper apparent logarithmic decay. Figure~\ref{fig:7}(\textit{c}) should therefore be interpreted primarily in terms of the qualitative trends with convection rather than a precise comparison with canonical $A_u$ values.

The fitted $B_u$ values in figure~\ref{fig:7}(\textit{d}) follow $B_u \approx 0.2 (-z_i/L)^{2/3} = 0.2w_\star^2/u_*^2$, indicating that the variance offset is predominantly governed by the convective velocity. The constant offset with respect to $w_\star$ ensures the logarithmic profile approaches a finite value independent of $u_*$ in the mixed layer where shear is negligible. The confinement of the buoyancy correction to the coefficient $B_u$ in the roll regime indicates that the impact of convection on near-surface $u'$ is non-local and approximately independent of height.

These results are qualitatively consistent with the extended attached-eddy formulation of \citet{Banerjee2015}, in which the logarithmic slope $A_u$ remains approximately constant relative to $u_*$ under weak convection while the intercept $B_u$ varies with stability. In the present LES, however, the approximately constant $A_u$ persists to stronger convection up to $-z_i/L\approx50$ and connects this range with the transition from rolls to cells. The transition point in $A_u$ corresponds to a local stability of $-\zeta\approx5$ at $z^\star=0.1$, compared to the predicted threshold of $-\zeta < 0.5$ and $z^\star < 0.02$ in \citet{Banerjee2015}. Additionally, the fitted intercept is seen here to depend on the global convection and $-z_i/L$ rather than the local $-\zeta$, suggesting that the outer mixed-layer dynamics exert an important influence on the surface-layer horizontal variance. The different $B_u$ dependence may also help to explain observed quantitative discrepancies between the \citet{Banerjee2015} predictions and other convective simulations \citep{Yang2020}.

Fitting the $\langle u'^2 \rangle$ profiles to the mean deficit formulation in table~\ref{tab:streamwise} yields similar values of $A_{u_2}$ and $B_{u_2}$ as \citet{Yang2020}. However, the profiles adhere more closely to a logarithmic decay than to the shape of the mean deficit, where the former leads to higher goodness-of-fit values. This difference is physically relevant because the logarithmic variance in figure~\ref{fig:7}(\textit{a}) and linear relation between $\langle u'^2\rangle$ and $U(z)$ in the mean-deficit formula implies the mean velocity profile is also logarithmic, but $U(z)$ departs substantially from the log law in the CBL, which corresponds to a height-dependent non-dimensional gradient $\phi_m$ \citep[see, e.g.,][]{Heisel2024}. While the mean deficit model operates well as a first-order approximation and yields consistent fitted constants, it requires further adjustments to account for the difference in the height dependence of $\langle u'^2 \rangle$ and $U$.

\begin{figure}
  \centering
  \includegraphics[width=1.0\textwidth]{figures/global_legend.eps}
  \vspace{0.2cm}
  \includegraphics[width=1.0\textwidth]{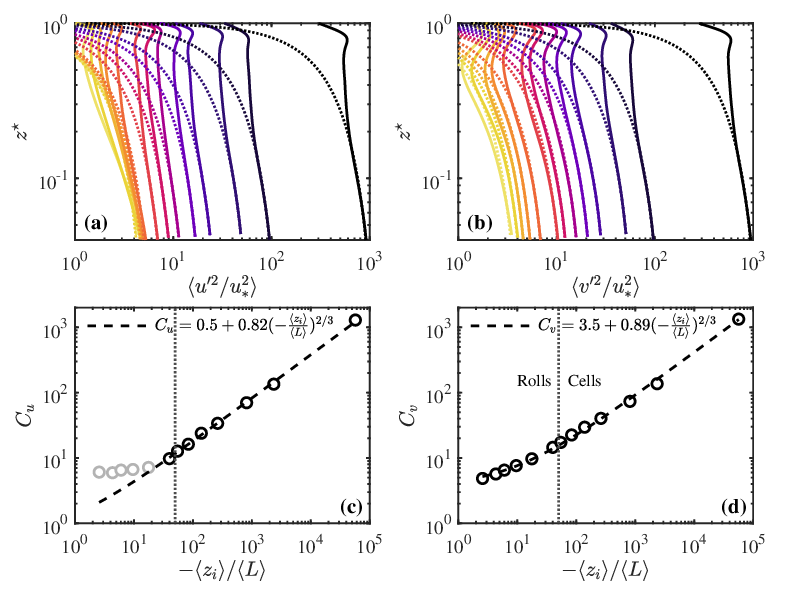}
  \caption{The horizontal velocity variances for strong convection fitted to $\langle u_i'^2 / u_*^2 \rangle =  C_i [1-(z^\star)^{0.38}]$ \citep{Wilson2008} within the surface layer. (\textit{a,b}) Profiles of the normalised streamwise and spanwise velocity variances,
  respectively, where the dotted lines represent surface layer fits. (\textit{c,d}) Fitted coefficients $C_u$ and $C_v$, respectively, as functions of global instability. In (\textit{c,d}), dashed lines represent
  empirical fits and grey markers identify the weak-convection streamwise cases whose profiles follow a logarithmic trend.}
  \label{fig:8}
\end{figure}

Under strong convection, the streamwise variance departs from the logarithmic profile evaluated in figure~\ref{fig:7} and appears to align more closely with the height dependence proposed by \citet{Wilson2008} shown in figure~\ref{fig:6}(\textit{a}). To assess this trend, figure~\ref{fig:8} compares the streamwise and spanwise variance profiles with $\langle u_i'^2 / u_*^2 \rangle =  C_i [1-(z^\star)^{0.38}]$ representing the height dependence from the \citet{Wilson2008} formulation in table~\ref{tab:streamwise}. The surface layer spanwise variance in figure~\ref{fig:8}(\textit{b}) is well described by the fitted height dependence across all cases, whereas the streamwise variance in figure~\ref{fig:8}(\textit{a}) only follows this dependence for cases c6 onward representing strong convection. The poor alignment with cases c1--c5 is more readily apparent from figure~\ref{fig:6}(\textit{a}). The fact that case c6 approximately follows both the logarithmic decay in figure~\ref{fig:7}(\textit{a}) and the power law in figure~\ref{fig:8}(\textit{a}) indicates a relatively smooth transition between the two dependencies. The consistency of the spanwise profile shape across cases and the transition in streamwise variance from \eqref{eq:2} to the same dependence as $v'$ coincides with the evolving large-scale structure: rolls predominantly impact $v'$ while $u'$ retains its shear-driven dependencies, and the more geometrically isotropic cells lead to greater similarity in $u'$ and $v'$. It is expected that the spanwise variance also becomes logarithmic under near-neutral conditions in the absence of rolls, but this cannot be assessed from the present cases.

The fitted coefficients in figure~\ref{fig:8}(\textit{c,d}) quantify how the variance levels change with global instability. The approximate invariance of cases c1--c5 in figure~\ref{fig:8}(\textit{c}) is consistent with the $u_*$ scaling of $A_u$ in figure~\ref{fig:7}(\textit{c}). The dashed curves in figure~\ref{fig:8}(\textit{c,d}) represent fits to the same empirical global instability dependence as \citet{Wilson2008} in table~\ref{tab:streamwise}. The fits lead to the horizontal variance profiles
\begin{equation}
  \frac{\langle u'^2 \rangle}{u_*^2} = \left[ 0.5 + 0.82
  \left({-}\frac{\langle z_i \rangle}{\langle L \rangle}\right)^{2/3} \right] \left[ 1-\left(z^\star\right)^{0.38} \right],
  \label{eq:Cu_wilson}
\end{equation}
\begin{equation}
  \frac{\langle v'^2 \rangle}{u_*^2} = \left[ 3.5 + 0.89
  \left({-}\frac{\langle z_i \rangle}{\langle L \rangle}\right)^{2/3} \right] \left[ 1-\left(z^\star\right)^{0.38} \right],
  \label{eq:Cv_wilson}
\end{equation}
where the exponents are imposed to match \citet{Wilson2008} but the coefficients differ. In particular, the $2/3$ exponent in \eqref{eq:Cu_wilson} and \eqref{eq:Cv_wilson} is consistent with the scaling $w_\star^2/u_*^2\propto(-z_i/L)^{2/3}$. The fitted trends are therefore consistent with the transition from surface shear to convective scaling with increasing buoyancy. The difference in offset values between $C_u$ and $C_v$ is required to account for the trends observed in figure~\ref{fig:4}, where the spanwise variance is greater in weak convection before the horizontal variances converge to the same value in free convection. A matching coefficient for the $-z_i/L$ term in $C_u$ and $C_v$ is expected to ensure this convergence, noting that the difference in the fitted values is within their combined 95\% statistical uncertainty ranges ($0.82 \pm 0.06$ for $u'$ and $0.89 \pm 0.06$ for $v'$).

\section{Variance budget statistics}
\label{sec:variancebudgets}

The preceding analyses identified statistical changes in the velocity variances from weak to free convection. The enhanced $v'$ and the non-monotonic evolution of $w'$ are suggestive of underlying trends in the production, transport, and redistribution of turbulent energy across the roll-to-cell transition. Therefore, the variance budget equations introduced in section~\ref{sec:budgets} are used here to assess the energy transfer pathways with changing convection.

\subsection{Pressure redistribution and return to isotropy}
\label{sec:redistribution_integrated_budgets}

\begin{figure}
  \centering
  \includegraphics[width=1.0\textwidth]{figures/global_legend.eps} 
  \vspace{0.2cm}
  \includegraphics[width=1.0\textwidth]{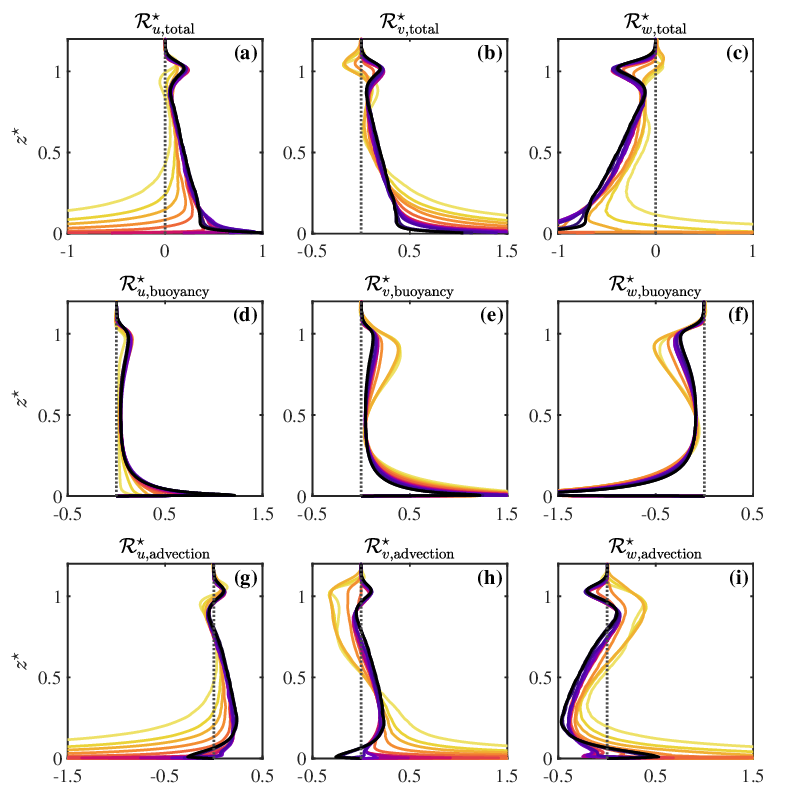}
  \caption{Profiles of the pressure-redistribution for the streamwise ($\mathcal{R}_u$, left), spanwise ($\mathcal{R}_v$, middle), and vertical ($\mathcal{R}_w$, right) velocities. Rows correspond to decomposition of the pressure field: (\textit{a}--\textit{c}) total redistribution $\mathcal{R}_{\mathrm{total}}$; (\textit{d}--\textit{f}) redistribution $\mathcal{R}_{\mathrm{buoyancy}}$ by pressure due to buoyancy; (\textit{g}--\textit{i}) redistribution $\mathcal{R}_{\mathrm{advection}}$ by pressure due to nonlinear advection.}
  \label{fig:9}
\end{figure}

The pressure redistribution $\mathcal{R}_i$ is the only mechanism in \eqref{eq:3}--\eqref{eq:5} for energy exchange across the velocity components. The redistribution profiles are shown in figure~\ref{fig:9}(\textit{a--c}), with the star superscript indicating local-in-time normalisation by $w_\star(t)^3/z_i(t)$ (see, e.g., section~\ref{sec:normalisation}). In the surface layer, shear production of $u'$ is redistributed to both $v'$ and $w'$ for the weakest convection case c1. For the remaining cases within the roll regime, $v'$ receives net energy from both $u'$ and $w'$. Energy distribution is from $w'$ to $u'$ and $v'$ in the surface layer under stronger convection and for all cases within the mixed layer.

The pressure field can be decomposed into contributions from individual terms in the Navier--Stokes equations by solving variations of the pressure Poisson equation based on each term \citep{Moeng1986,Heinze2016,Ding2018,Yi2025}. Figure~\ref{fig:9}(\textit{d--f}) shows $\mathcal{R}_{i,\mathrm{buoyancy}}$ due to pressure fluctuations associated with buoyancy arising from the gravity force term and the Boussinesq approximation. Similarly, figure~\ref{fig:9}(\textit{g--i}) shows $\mathcal{R}_{i,\mathrm{advection}}$ due to the pressure field from the nonlinear advection term. Advection can be further decomposed into shear and turbulent-turbulent components \citep{Heinze2016}, but the turbulent-turbulent contribution was found to be dominant with no unique convection trends revealed by the additional decomposition. Pressure statistics due to other terms such as Coriolis are negligibly small and not shown here.

The buoyancy contribution to the vertical component, $\mathcal{R}^{\star}_{w,\mathrm{buoyancy}}$ in figure~\ref{fig:9}(\textit{f}), transfers buoyancy-generated energy away from $w'$ toward the horizontal velocity components for all heights and cases. Its profiles collapse across the strongly convective cases when normalised by $w_\star$, with all profiles exhibiting a similar vertical distribution. The minimum value in the mixed layer and local peaks at the top and bottom of the CBL correspond to the large-scale rolls and cells, in which the vertical motions in the mid-CBL connect to horizontal displacements at the boundaries. The horizontal redistributions, $\mathcal{R}^{\star}_{u,\mathrm{buoyancy}}$ and $\mathcal{R}^{\star}_{v,\mathrm{buoyancy}}$ in figure~\ref{fig:9}(\textit{d,e}), have the same trend as $w'$ but with positive $\mathcal{R}_i$. The strong near-surface peaks in figure~\ref{fig:9}(\textit{d--f}) reflect the kinematic wall-blocking pressure response enforced by the wall boundary condition \citep{Glendening2000, Hunt2000, ChavezMedina2025}.  Under weak convection, most of the buoyancy-related redistribution at the top and bottom of the CBL goes to $\mathcal{R}^{\star}_{v,\mathrm{buoyancy}}$ rather than $\mathcal{R}^{\star}_{u,\mathrm{buoyancy}}$, consistent with the circulations of the longitudinal rolls. The difference between the two progressively decreases across the rest of the parameter space until the redistribution to $u'$ and $v'$ is equal for cells under free convection.

The advection contribution exhibits substantially greater variation with height and convection. The redistribution of shear-generated turbulence from $\mathcal{R}^{\star}_{u,\mathrm{advection}}$ in figure~\ref{fig:9}(\textit{g}) to $\mathcal{R}^{\star}_{v,\mathrm{advection}}$ and $\mathcal{R}^{\star}_{w,\mathrm{advection}}$ in figure~\ref{fig:9}(\textit{h,i}) decreases with diminishing shear, and the profiles become approximately invariant for cases c7--c13. The non-zero profiles in these cases indicate an indirect contribution of buoyancy through the turbulent-turbulent advection term, such that the trends in figure~\ref{fig:9}(\textit{g--i}) do not solely reflect shear effects.

\begin{figure}
  \centering
  \includegraphics[width=1.0\textwidth]{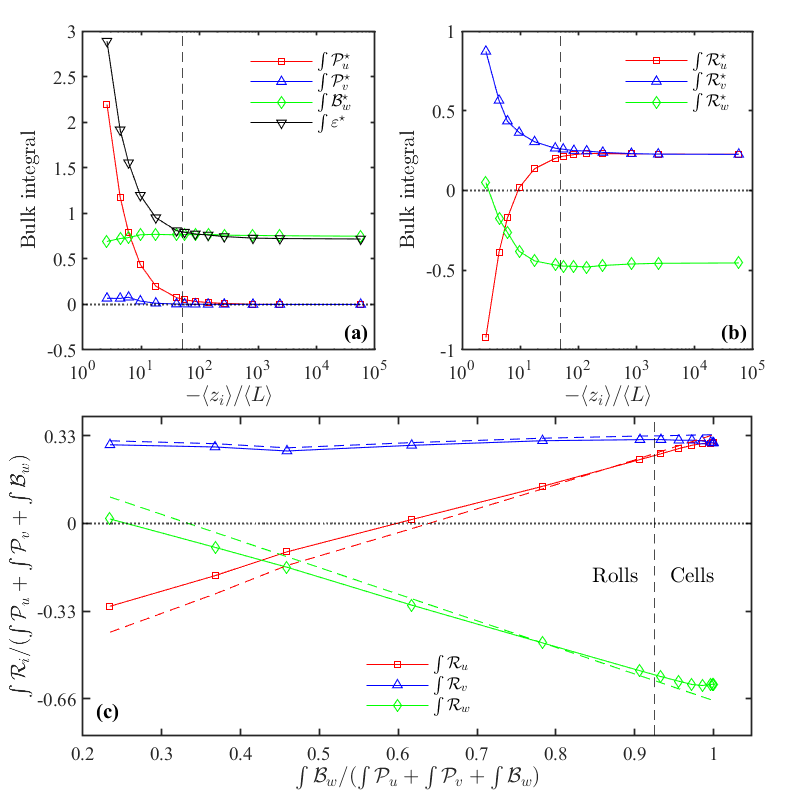}
  \caption{Variance budget statistics integrated across the CBL depth. (\textit{a}) Production of turbulence by shear ($\textstyle\int \mathcal{P}_u^\star$ and $\textstyle\int \mathcal{P}_v^\star$) and buoyancy ($\textstyle\int \mathcal{B}_w^\star$) as well as dissipation ($\textstyle\int \varepsilon^\star$) as functions of global instability $-z_i/L$. (\textit{b}) Redistribution $\textstyle\int \mathcal{R}_i^\star$ across velocity components by pressure--strain correlations versus $-z_i/L$; (\textit{c}) Redistribution relative to the total rate of production versus the fraction of production due to buoyancy. The vertical dashed line across all panels denotes the approximate end of the transition from roll- to cell-dominated convection at $-z_i/L \approx 50$. The coloured dashed lines in (\textit{c}) represent the \eqref{eq:7} return-to-isotropy predictions for $\mathcal{R}_i$ leading to isotropic dissipation.}
  \label{fig:10}
\end{figure}

To further evaluate the trends with varying convection, figure~\ref{fig:10} shows the depth-integrated redistribution terms in the same manner as the variances in figure~\ref{fig:5}. Also included are the production and dissipation budget terms in figure~\ref{fig:10}(\textit{a}). The integrated statistics contain uncertainty associated with the estimation of the SGS contributions and with the absence of information between the wall and the first grid point in the wall-modelled LES. Accordingly, the interpretation of figure~\ref{fig:10} emphasises the trends across the parameter space rather than the absolute magnitudes of the integrated quantities.

The productions in figure~\ref{fig:10}(\textit{a}) show a clear transition in the dominant source of turbulent energy. The integrated streamwise shear production, $\int\mathcal{P}_u^\star$, decreases rapidly with increasing instability and becomes small relative to buoyancy by the end of the transition from rolls to cells. The spanwise shear production $\int\mathcal{P}_v^\star$ remains close to zero throughout the parameter range, with the small values under weak convection corresponding to production in the entrainment layer where there is strong wind shear and veering as seen in figure~\ref{fig:1}. The integrated buoyancy production, $\int\mathcal{B}_w^\star$, remains relatively constant across cases. By definition, the normalised production $\mathcal{B}_w z_i/w_\star^3$ reduces to the heat flux profile $\langle w'\theta'\rangle / Q$, see, e.g., figure~\ref{fig:3}(\textit{b}), which is invariant with convection aside from small changes in the negative flux at $z_i$. Buoyancy production becomes the dominant source of TKE as shear diminishes, with more than $90\%$ of the total integrated production due to buoyancy at $-z_i/L=50$. The integrated dissipation decreases with the decreasing shear and approximately balances the buoyancy production under the strongly convective cases, consistent with the quasi-steady state of the flow.

The evolution of the pressure-redistribution terms shown in figure~\ref{fig:10}(\textit{b}) is consistent with the trends observed in the figure~\ref{fig:9} profiles. In the weakly convective roll regime, there is a net transfer of energy away from the streamwise component $\int\mathcal{R}_u^\star$ to the spanwise component $\int\mathcal{R}_v^\star$ and to a lesser extent the vertical component $\int\mathcal{R}_w^\star$. Energy redistribution away from $w'$ progressively increases with convection until reaching an approximate constant in the cell regime. Given the contribution of buoyancy to $\mathcal{R}_w^\star$ is fixed relative to the convective scales (figure~\ref{fig:9}(\textit{f})), this trend is due to a reduction in redistribution from $u'$ to $w'$ as surface shear diminishes to negligible values (figures~\ref{fig:9}(\textit{g,i}) and \ref{fig:10}(\textit{a})). The horizontal redistributions $\int\mathcal{R}_u^\star$ and $\int\mathcal{R}_v^\star$ converge to similar values in the same manner as figure~\ref{fig:9}(\textit{d,e}), corresponding to the progressive loss of horizontal alignment in the roll structures as they transition toward isotropic cellular plumes.

In addition to its connection to the large-scale structure of the CBL, pressure redistribution is often considered for its role in isotropisation of the flow \citep{BouZeid2018}, noting a common feature of pressure closure models is a return-to-isotropy component \citep{Pope2000}. In the present context, isotropy refers to the dissipation tensor and its components in \eqref{eq:3}--\eqref{eq:5}. The vertical transport terms in these budget equations are zero upon integration across the CBL depth, given that transport only acts to distribute existing energy across heights. The Coriolis contribution is negligible and the storage term associated with the slow growth of the CBL is small relative to the remaining terms, such that there is an approximate balance between production, redistribution, and dissipation in the integrated budgets:
\begin{align}\label{eq:6}
0 &\approx \int \mathcal{P}_u + \int \mathcal{R}_u - \int \varepsilon_u, \nonumber\\
0 &\approx \int \mathcal{P}_v + \int \mathcal{R}_v - \int \varepsilon_v, \\
0 &\approx \int \mathcal{B}_w + \int \mathcal{R}_w - \int \varepsilon_w. \nonumber
\end{align}
Same as before, the integral notation is shorthand for the integral across depth, e.g. $\int_0^\infty \mathcal{P}_u\,\mathrm{d}z$. Dividing through by the total production and assuming the dissipation is isotropic and in balance with the total production leads to the following return-to-isotropy prediction for the redistribution:
\begin{align}\label{eq:7}
\frac{\int \mathcal{R}_u}{\int \mathcal{P}_u + \int \mathcal{P}_v + \int \mathcal{B}_w} &\approx \frac{1}{3} - \frac{\int \mathcal{P}_u}{\int \mathcal{P}_u + \int \mathcal{P}_v + \int \mathcal{B}_w}, \nonumber\\
\frac{\int \mathcal{R}_v}{\int \mathcal{P}_u + \int \mathcal{P}_v + \int \mathcal{B}_w} &\approx \frac{1}{3} - \frac{\int \mathcal{P}_v}{\int \mathcal{P}_u + \int \mathcal{P}_v + \int \mathcal{B}_w}, \\
\frac{\int \mathcal{R}_w}{\int \mathcal{P}_u + \int \mathcal{P}_v + \int \mathcal{B}_w} &\approx \frac{1}{3} - \frac{\int \mathcal{B}_w}{\int \mathcal{P}_u + \int \mathcal{P}_v + \int \mathcal{B}_w}. \nonumber
\end{align}
These equations match \citet{BouZeid2018} but with $\mathcal{B}_w$ included in the normalisation. The integrated LES redistributions are compared with these predictions in figure~\ref{fig:10}(\textit{c}) as the fraction of energy production due to buoyancy varies from 0 to 1, noting that the present work does not include fractions less than 0.2 which encompasses the near-neutral regime.

The redistributions in figure~\ref{fig:10}(\textit{c}) align relatively well with the \eqref{eq:7} return-to-isotropy predictions, including the approach to $+1/3$ for each horizontal component and $-2/3$ for the vertical component as buoyancy becomes the sole source of TKE under free convection. The modest discrepancies from the prediction may in part be due to missing SGS contributions and statistics below the first grid point, which is particularly relevant for the surface production and redistribution under weak convection. Accordingly, the primary conclusion from figure~\ref{fig:10}(\textit{c}) is that the bulk redistribution of energy across the CBL is consistent with the isotropisation of $\varepsilon$ to first-order accuracy.

Because $\int\mathcal{R}_v$ does not exceed the return-to-isotropy prediction, the redistribution magnitude in figure~\ref{fig:10} cannot by itself explain how $v'$ is the largest turbulent component under certain convective conditions as seen in figure~\ref{fig:5}. In the following section, the spectrum of the variance budget is used to determine whether scale-wise trends can elucidate how the global isotropisation of dissipation leads to the observed variance statistics.

\subsection{Spectral variance budgets and scale-dependent energy transfer}
\label{sec:spectral_budgets}

The variance budgets in \eqref{eq:3}--\eqref{eq:5} are transformed into the wavenumber domain to identify the characteristic horizontal scales and heights at which production, pressure transport, and pressure--strain redistribution occur \citep{Lee2019,Huang2026}. An additional scale-to-scale energy transfer term arises in the wavenumber domain \citep{Pope2000}. This new term is not discussed here, except to note that the net scale-to-scale transfer is predominantly downscale across the sampled heights and convection states in the present simulations. The spectral analysis is used here to identify the dominant scales and pathways of variance production, transport, and redistribution.

The co-spectrum of each budget term was evaluated from two-dimensional (2D) Fourier transforms of the horizontal fields from 60 instantaneous 3D flow volumes distributed throughout each simulation. As an example, the spectral pressure-transport term $\widehat{\Pi}_w^\star$ is estimated from the co-spectrum of $w'$ and $p'$, with the vertical derivative applied to the resulting co-spectrum. The other spectral budget terms were obtained analogously from the velocity, pressure, and temperature co-spectra. Here, the hat notation $\widehat{(\cdot)}$ denotes the spectral density of a given variable. The 2D spectral densities across wavenumbers $k_x$ and $k_y$ were translated to 1D functions by integrating the spectra azimuthally along fixed values of the horizontal wavenumber magnitude $k = (k_x^2+k_y^2)^{1/2}$. Same as the previous results, the star (${}^\star$) superscript denotes local-in-time normalisation by $w_\star^3(t)$ for the spectra and $z_i(t)$ for the wavenumber. The spectral densities were interpolated to common values of both height $z/z_i$ and wavenumber magnitude $k z_i$ prior to averaging across flow volumes.

\begin{figure}
  \centering
  \includegraphics[width=1.0\textwidth]{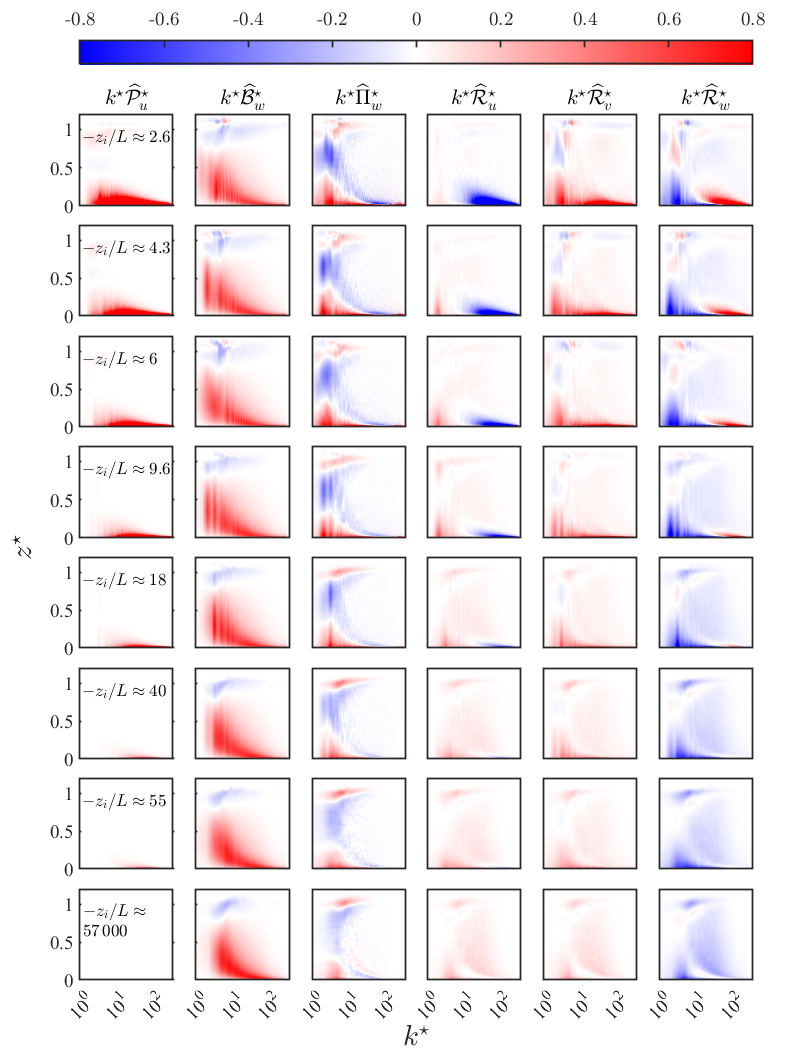}
  \caption{Premultiplied colour spectra of variance budget terms as functions of the normalised wavenumber $k^\star = k z_i$ and height $z^\star$. Rows correspond to cases c1--c7 spanning the transition from rolls to cells and free convection case c13. Columns show the streamwise shear production, $\widehat{\mathcal{P}}_u^\star$, vertical buoyancy production, $\widehat{\mathcal{B}}_w^\star$, vertical pressure transport, $\widehat{\Pi}_w^\star$, and the pressure-–strain redistribution terms $\widehat{\mathcal{R}}_u^\star$, $\widehat{\mathcal{R}}_v^\star$, and $\widehat{\mathcal{R}}_w^\star$.}
  \label{fig:11}
\end{figure}

Colour contours of the premultiplied variance budget spectral densities are shown in figure~\ref{fig:11} to visualise the scale-dependent pathways of turbulent energy production and redistribution.  The changes are gradual within the cell regime, such that only the limits of this regime (c7 and c13) are shown here. Under weak convection within the roll regime, the spectral budgets identify two distinct pathways transferring energy to the spanwise velocity. Near the surface, streamwise variance produced by mean shear $\widehat{\mathcal{P}}_u$ across a range of scales is locally redistributed through pressure--strain interactions $\widehat{\mathcal{R}}_u$, with a preferential contribution to the spanwise component $\widehat{\mathcal{R}}_v$. Separately, buoyancy production $\widehat{\mathcal{B}}_w$ in the mixed layer generates vertical turbulence at moderately larger scales, which is subsequently transported downward by pressure $\widehat{\Pi}_w$ and redistributed near the surface through $\widehat{\mathcal{R}}_w$ to $\widehat{\mathcal{R}}_v$.

This vertical transport and displacement from $w'$ to $v'$ corresponds to the large-scale roll motions, indicating the thermal rolls are mediated by pressure. Large eddies approaching the wall experience kinematic wall-blocking \citep{Glendening2000,Hunt2000}, which suppresses wall-normal motion and requires a compensating horizontal response to satisfy continuity \citep{Hunt2000,Nguyen2013,Jayaraman2021}. This process occurs through the wall-reflection pressure, which increases the vertical pressure gradient near the surface and redistributes part of the vertical turbulent energy into horizontal motions \citep{Pope2000,Ding2018,ChavezMedina2025}. This redistribution from the vertical component is preferentially directed toward $v'$ due to the alignment of the thermal rolls.

As the large-scale flow organisation restructures from rolls to cells, the spectral energy pathways also change. The surface shear production progressively diminishes across cases c1{--}c6, leading to negligible redistribution from $\widehat{\mathcal{R}}_u$ to $\widehat{\mathcal{R}}_v$ by $-z_i/L \approx 40\text{--}55$. At the same time, the largest scales of buoyancy production $\widehat{\mathcal{B}}_w$ occur at moderately higher wavenumbers under stronger convection. This shift towards smaller scales is attributed to both a loss of streamwise elongation of the structures and a concurrent reduction in the characteristic lateral spacing of the cellular plumes relative to the rolls \citep{Kaimal1976, Khanna1998, Salesky2018}. Pressure--strain redistribution $\widehat{\mathcal{R}}_w$ to the two horizontal components becomes increasingly equal with increasing convection and $\widehat{\mathcal{R}}_u$ and $\widehat{\mathcal{R}}_v$ appear to be the same under free convection. This convergence is consistent with figure~\ref{fig:10}(\textit{b}) and the loss of directional preference in the structures as shear reduces.

The spectra in figure~\ref{fig:11} further reveal that the large-scale, pressure-mediated coupling between the mixed layer and the surface weakens with increasing convection. Both the pressure transport $\widehat{\Pi}_w$ connecting these layers and the near-surface redistribution $\widehat{\mathcal{R}}_w$ decrease with increasing convection across the parameter space. The results indicate that thermal rolls more efficiently transfer large-scale energy across heights and velocity components compared to cells. At the same time, redistribution from $w'$ to the horizontal velocities increases across a range of smaller scales at all heights throughout the CBL, leading to a preservation of the net redistribution seen in figure~\ref{fig:9}(\textit{c}).

\begin{figure}
  \centering
  \includegraphics[width=1.0\textwidth]{figures/global_legend.eps}
  \vspace{0.2cm}
  \includegraphics[width=1.0\textwidth]{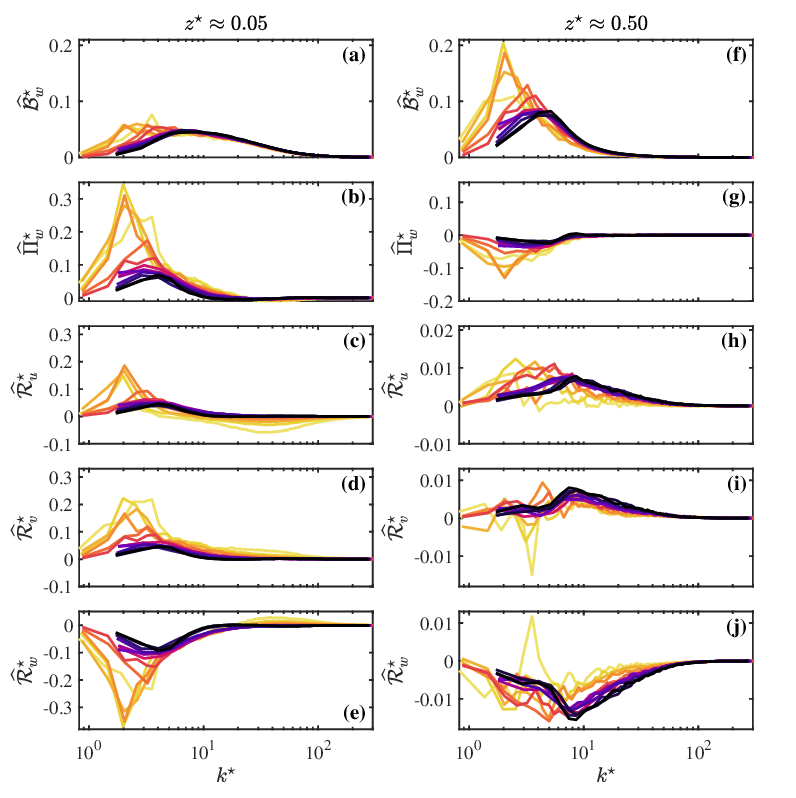}
  \caption{Spectra of the variance budget terms as functions of the normalised wavenumber, $k^\star$, for the near-surface ($z^\star\approx0.05$; \textit{a}--\textit{e}) and mixed-layer ($z^\star\approx0.50$; \textit{f}--\textit{j}) regions. The rows show vertical buoyancy production, $\widehat{\mathcal{B}}_w^\star$, vertical pressure transport, $\widehat{\Pi}_w^\star$, and pressure--strain redistribution for the streamwise, spanwise, and vertical velocity variances, $\widehat{\mathcal{R}}_u^\star$, $\widehat{\mathcal{R}}_v^\star$, and $\widehat{\mathcal{R}}_w^\star$, respectively.}
  \label{fig:12}
\end{figure}

The changing magnitudes of the buoyancy-driven energy pathways observed in figure~\ref{fig:11} are further assessed in figure~\ref{fig:12} which shows the scale-wise budget terms for all cases at two representative heights to evaluate the changes in low-wavenumber statistics. Figure~\ref{fig:12}(\textit{a--e}) shows the near-surface spectra ($z^\star \approx 0.05$), while figure~\ref{fig:12}(\textit{f--j}) shows the corresponding mixed-layer spectra ($z^\star \approx 0.50$). The spike at $k^\star \approx 3\text{--}4$ for case c1 in figure~\ref{fig:12}(\textit{i,j}) corresponds to a negative $\widehat{\mathcal{R}}_v^\star$ region in the upper half of the CBL seen in figure~\ref{fig:11}, which is attributed to top-down effects related to entrainment layer production which is outside the scope of the present analysis.

Net buoyancy production is largest at the surface in figure~\ref{fig:12}(\textit{a}), but near-surface production is relatively broadband. There is a significantly higher narrowband concentration of large-scale $\widehat{\mathcal{B}}_w^\star$ in the mixed layer (figure~\ref{fig:12}(\textit{f})) despite overall lower net production farther from the surface. For the weak convection cases, a significant portion of this large-scale production is transported by pressure from the mixed layer (figure~\ref{fig:12}(\textit{g})) towards the surface (figure~\ref{fig:12}(\textit{b})). Almost all of this energy is transferred to the horizontal velocity components (figure~\ref{fig:12}(\textit{c,d})) with preferential redistribution to $v'$, noting $\widehat{\Pi}_w^\star$ in figure~\ref{fig:12}(\textit{b}) and $\widehat{\mathcal{R}}_w^\star$ in figure~\ref{fig:12}(\textit{e}) have approximately equal magnitudes in the low-wavenumber range. The redistribution of shear-generated turbulence from $u'$ to $v'$ and $w'$ (figure~\ref{fig:12}(\textit{d,e})) is also apparent at smaller scales ($k^\star \gtrsim 10$) under weak convection. While this combined redistribution of both shear- and buoyancy-generated turbulence to $v'$ does not exceed isotropy predictions, the redistribution mediated by pressure and the roll motions originates in the mixed layer and occurs at the largest scales in the flow.

Consistent with figure~\ref{fig:11}, the concentrated large-scale production, transport, and redistribution peaks are progressively lost in figure~\ref{fig:12} as the rolls (c1) evolve into cells (c7 onward) with increasing convection. At the same time, the weakening peaks shift toward higher normalised wavenumber from $k^\star \approx 2$ to $k^\star \approx 4\text{--}5$ in figure~\ref{fig:12}(\textit{b--f}). In the mixed layer, the weaker pressure transport toward the surface for cases c7--c13 in figure~\ref{fig:12}(\textit{g}) is compensated by increased pressure--strain redistribution in figure~\ref{fig:12}(\textit{j}). However, this increase is specifically at smaller scales ($k^\star \gtrsim 5$), and there is a deficit in $\widehat{\mathcal{R}}_w^\star$ relative to the roll cases at the scale of the  production peaks ($k^\star \approx 4\text{--}5$ in figure~\ref{fig:12}(\textit{f})). The deficit indicates that more large-scale buoyancy-generated energy is retained locally within $w'$ with subsequent downscale transfer and redistribution at smaller scales, thus providing a possible explanation for the increase in the $w'$ peak across the cell regime in figure~\ref{fig:4}(\textit{c}). The downward displacement of the peak toward $z^\star\approx0.3$ may be related to a coinciding shift in the symmetry of the pressure transport; the height with low-wavenumber $\widehat{\Pi}_w^\star=0$ in figure~\ref{fig:11} moves downward with increasing convection in a manner consistent with the $w'$ peak. However, this interpretation remains speculative and requires further investigation.

\begin{figure}
  \centering
  \includegraphics[width=1.0\textwidth]{figures/global_legend.eps}
  \vspace{0.2cm}
  \includegraphics[width=1.0\textwidth]{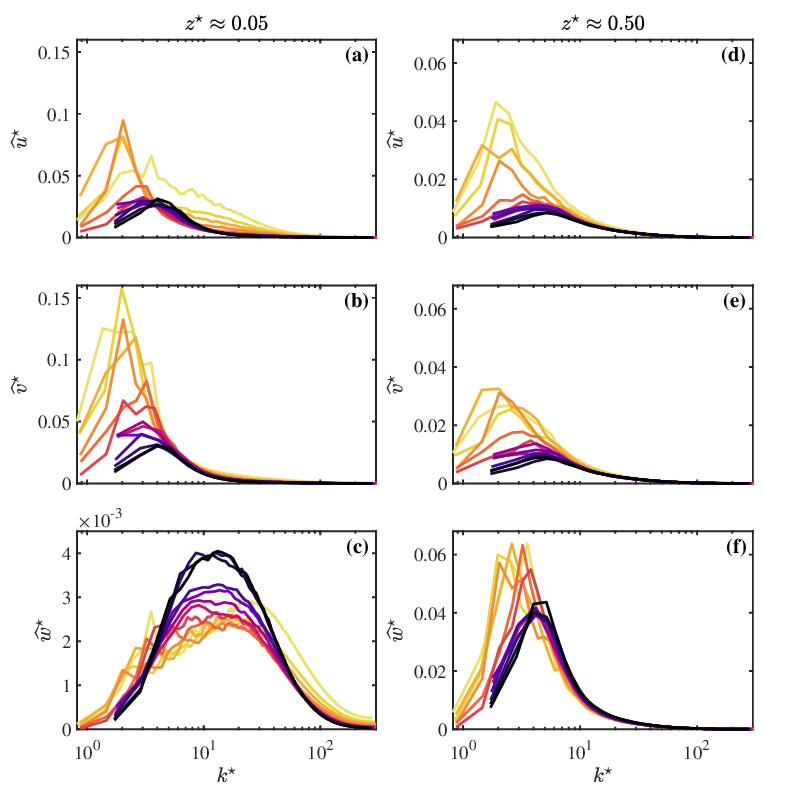}
  \caption{Velocity spectra as functions of the normalised wavenumber, $k^\star$, for the near-surface ($z^\star\approx0.05$; \textit{a}--\textit{c}) and mixed-layer ($z^\star\approx0.50$; \textit{d}--\textit{f}) regions. The rows show the streamwise $\widehat{u}^\star$, spanwise $\widehat{v}^\star$, and vertical $\widehat{w}^\star$ velocity spectra, respectively. The same ordinate range is plotted for each component at a given height, except for (\textit{c}) due to its substantially smaller magnitude.}
  \label{fig:13}
\end{figure}

Figure~\ref{fig:13} provides the 1D velocity variance spectral densities for the same heights as figure~\ref{fig:12} to facilitate discussion on how the energy pathways identified from figures~\ref{fig:11} and \ref{fig:12} result in the variance trends observed in section~\ref{sec:variance}. The 1D velocity spectra are denoted by $\widehat{u}^\star$, $\widehat{v}^\star$, and $\widehat{w}^\star$ for the streamwise, spanwise, and vertical components, respectively, where the star (${}^\star$) superscript denotes local-in-time normalisation by $w_\star^2(t)z_i(t)$. 

Near the surface, the roll-dominated cases exhibit a distinct low-wavenumber peak in both horizontal components in figure~\ref{fig:13}(\textit{a,b}) at $k^\star\approx2$, the same wavenumber as the large-scale buoyancy production, pressure transport and redistribution peaks in figure~\ref{fig:12}. The spanwise $\widehat{v}^\star$ low-wavenumber peak exceeds the streamwise $\widehat{u}^\star$ peak across the roll regime in cases c1--c6, consistent with the preferential redistribution toward $v'$ identified from $\widehat{\mathcal{R}}_v^\star$ in figure~\ref{fig:12}. While the overall redistribution $\textstyle\int\mathcal{R}$ approximately aligns with return-to-isotropy (figure~\ref{fig:10}(\textit{c})), the energy redistributed to $v'$ by roll motions is predominantly at the largest scales (figure~\ref{fig:12}(\textit{d})), whereas dissipation $\varepsilon$ is exclusively within subgrid scales in the present LES. This pathway allows for $v'$ to accumulate more large-scale energy than the other velocity components, leading to larger variance statistics.

In the comparison of horizontal variances, the greater large-scale energy in $\widehat{v}^\star$ is partially offset by higher $\widehat{u}^\star$ at intermediate and smaller scales in cases c1--c6. While some of this shear-generated energy is redistributed from $u'$ to $v'$ (figures~\ref{fig:9}(\textit{h}) and \ref{fig:12}(\textit{d})), the greater small-scale energy in $\widehat{u}^\star$ reduces the difference between components in figure~\ref{fig:5}. In contrast, the vertical velocity spectrum $\widehat{w}^\star$ near the surface in figure~\ref{fig:13}(\textit{c}) has smaller magnitudes centred at higher wavenumbers, consistent with the redistribution of large-scale buoyancy-generated energy into the horizontal components (figure~\ref{fig:12}(\textit{e})).

The velocity spectra under stronger convection are also consistent with the previous figures: the spectral peaks shift to moderately higher wavenumbers as the large-scale structures lose their streamwise elongation and contract in lateral scale. Further, the horizontal velocity spectra become increasingly similar to each other at both heights, consistent with the increasingly symmetric pressure--strain redistribution between $u'$ and $v'$ seen in figure~\ref{fig:12}. The increase in the $w'$ peak with convection is less evident in figure~\ref{fig:13}(\textit{f}), although a modest increasing trend is apparent in the wavenumber range $k^\star=5\text{--}10$. One consideration is that the downward shift in the peak leads to a maximum variance at different heights across cases, such that the spectra at a fixed height may partially obscure spectral trends associated with the peak.

\section{Summary}
\label{sec:conclusions}

The present study assesses turbulent velocity variances in the CBL using thirteen wall-modelled LES cases summarised in table~\ref{tab:cases}. Each case exhibits a pronounced mixed layer (figure~\ref{fig:1}), such that near-neutral conditions are not considered here. The LES spans the remainder of the sheared convection parameter space, including six cases within the transition from large-scale longitudinal rolls to polygonal cells ending around $-z_i/L\approx50$, followed by seven cases within the cell regime as the flow asymptotically approaches penetrative free convection (figure~\ref{fig:2}).

This structural transition is accompanied by systematic changes in the velocity variances apparent from both the profiles (figure~\ref{fig:4}) and depth-integrated statistics (figure~\ref{fig:5}). The horizontal variances and TKE progressively decrease in magnitude relative to $w_\star$. The spanwise (cross-wind) variance exceeds the streamwise component for almost all of the cases and is the most energetic velocity component for weak-to-moderate convection (i.e. $-z_i/L\approx4\text{--}10$). The vertical (wall-normal) variance yields better similarity with $w_\star$ compared to the horizontal components (figure~\ref{fig:4}(\textit{c})) and agrees well with existing relations \citep{Lenschow1980}, but the magnitude of the $w'$ peak exhibits a weak non-monotonic trend with convection and its position moves downward approaching free convection.

Within the surface layer, the horizontal variances are poorly captured by MOST due to a height dependence separate from the local buoyancy effects represented by $-z/L$ (figure~\ref{fig:6}(\textit{a,b})). In contrast, the vertical variance is well described by MOST (figure~\ref{fig:6}(\textit{c})), particularly with local-in-height scaling (figure~\ref{fig:6}(\textit{d})), demonstrating that the vertical velocity responds more directly to the local balance of shear and buoyancy than do the horizontal velocity components.

Under weaker convection within the surface layer, corresponding to cases c1--c6 and the presence of longitudinal rolls, the streamwise velocity variance retains the logarithmic dependence in \eqref{eq:2} applicable to neutral conditions  (figure~\ref{fig:7}). The fitted logarithmic slope $A_u$ remains approximately constant with respect to $u_*$ consistent with previous predictions for this regime \citep{Banerjee2015}, while the fitted offset $B_u$ is proportional to $w_\star$. The latter enforces the mixed layer boundary condition above the surface layer, for which $w_\star$ is the leading-order scaling parameter. For cells under strong convection, both horizontal variances more closely follow the empirical formulation of \citet{Wilson2008} which includes $z_i/L$ and $z/z_i$ dependencies (figure~\ref{fig:8}). These results indicate that the height dependence of the streamwise variance remains predominantly organised by mechanical shear in the presence of coherent rolls which have spanwise-aligned circulations (figure~\ref{fig:13}(\textit{b})), whereas the horizontal variances become increasingly similar to each other as the preferential orientation of the rolls is lost (figure~\ref{fig:9}(\textit{d,e})).

The trends in the variance statistics are accompanied by analysis of the variance budget contributions in \eqref{eq:3}--\eqref{eq:5}. Streamwise shear production decreases rapidly with increasing convection and supplies less than 10\% of the total production across the CBL by the conclusion of the roll-to-cell transition (figure~\ref{fig:10}(\textit{a})). As energy production changes, the overall pressure--strain redistribution adjusts to promote isotropic dissipation (i.e. return to isotropy) to first-order accuracy (figure~\ref{fig:10}(\textit{c})).

\begin{figure}
  \centering
  \includegraphics[width=1.0\textwidth]{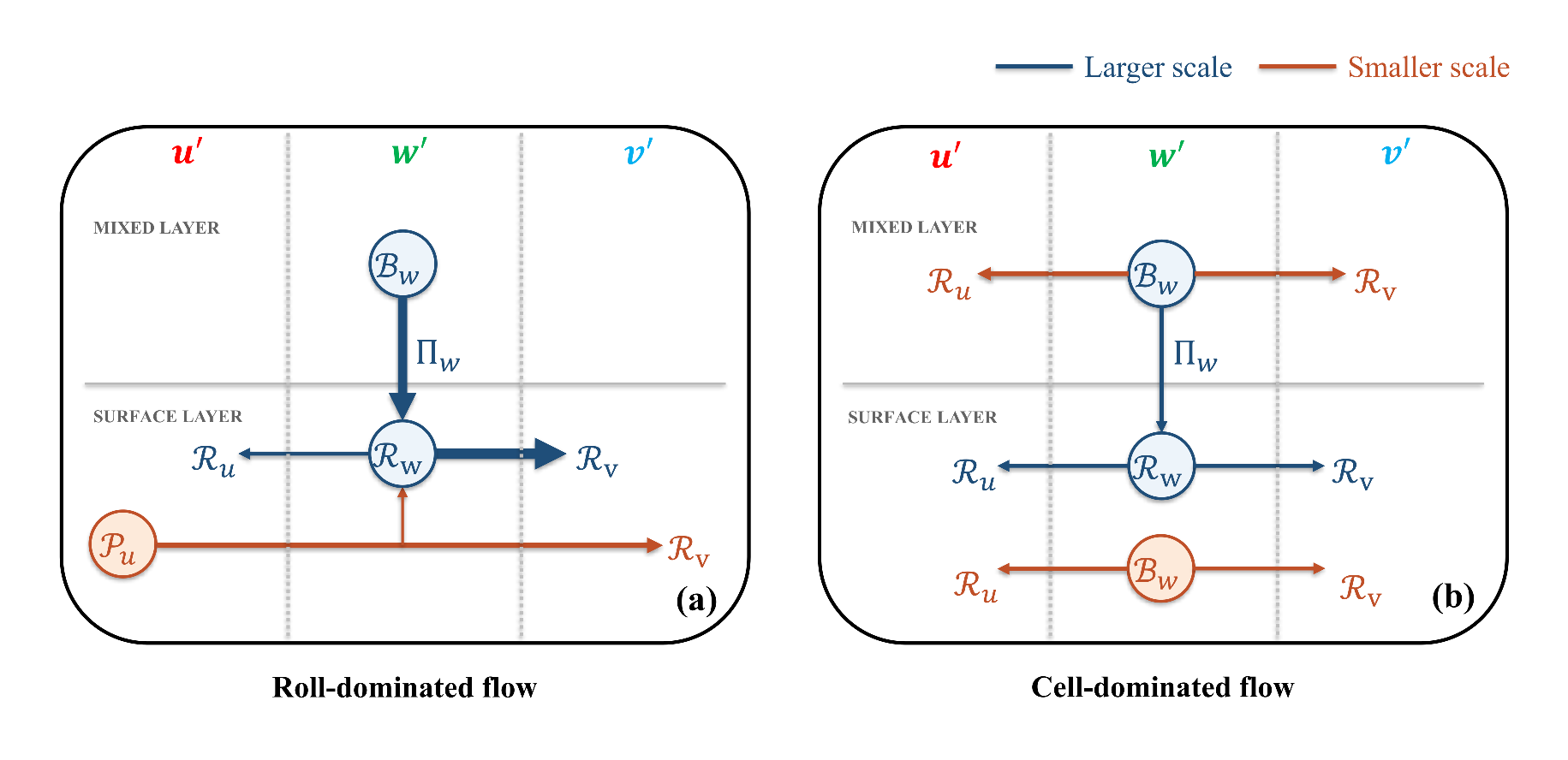}
  \caption{Schematic summary of the dominant pathways for turbulent energy production, transport, and redistribution in the CBL: for the (\textit{a}) roll-dominated and (\textit{b}) cell-dominated regimes. Blue and orange arrows denote predominantly larger- and smaller-scale pathways, respectively. Arrow thickness denotes the relative strength or preferential nature of the corresponding transfer.}
  \label{fig:14}
\end{figure}

Within the thermal roll regime represented by figure~\ref{fig:14}(\textit{a}), the largest-scale energy is produced by buoyancy in the mixed layer before being transported to the surface and redistributed preferentially to $v'$ by pressure. In the transition to cells represented by figure~\ref{fig:14}(\textit{b}), the large-scale pressure transport connecting the surface and mixed layers progressively reduces, and the redistribution between the horizontal components becomes more even (figures~\ref{fig:11} and \ref{fig:12}). As this pathway weakens, the redistribution $\mathcal{R}$ becomes more active at smaller scales across all heights (figure~\ref{fig:11}). These trends allow for increased local accumulation of large-scale $w'$ in the mixed layer, helping to explain the increasing vertical variance peak under strong convection. Further, there is a shift in the largest scales towards moderately higher wavenumbers as the rolls lose their streamwise coherence and transition to cells of smaller horizontal extent. The continuous decrease in TKE across the cases is attributed to this reduction in the largest scales as the cells contract and become geometrically isotropic. In the context of figure~\ref{fig:14}, the roll-to-cell transition can therefore be interpreted as a reorganisation of the buoyancy-driven motions from large-scale, pressure-mediated rolls connecting the surface and mixed layers to more local, smaller-scale cells with increasingly symmetric horizontal redistribution.

\begin{bmhead}[Acknowledgements.]
The authors acknowledge the high-performance computing resources provided by the Sydney Informatics Hub, a Core Research Facility of the University of Sydney, and the National Computational Infrastructure (NCI Australia), which is supported by the Australian Government.
\end{bmhead}

\begin{bmhead}[Funding.]
This research received no specific grant from any funding agency, commercial or not-for-profit sectors.
\end{bmhead}

\begin{bmhead}[Declaration of interests.]
The authors report no conflict of interest.
\end{bmhead}

\begin{appen}
\renewcommand{\theHsection}{\Alph{section}}

\section{Computational domain sensitivity}
\label{sec:appA}

\begin{figure}
  \centering
  \includegraphics[width=1.0\textwidth]{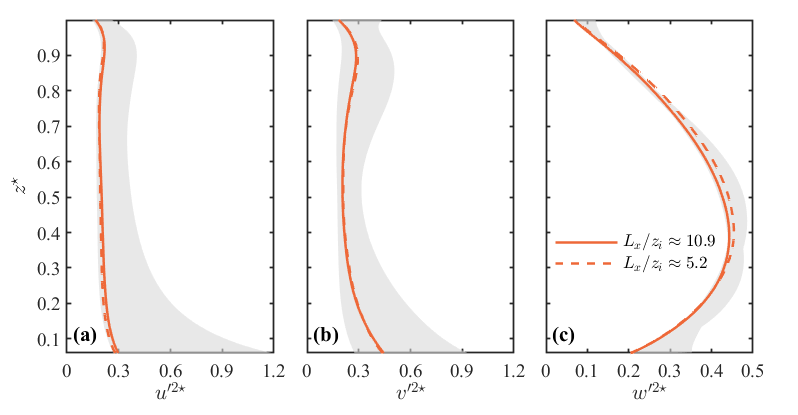}
  \caption{Comparison of moderate-convection case c6 statistics using the larger (solid lines) and smaller (dashed lines) horizontal domains from table~\ref{tab:cases}, where the legend indicates the domain length $L_x$ relative to $z_i$. Normalised streamwise, spanwise, and vertical velocity variance profiles are shown in (\textit{a}), (\textit{b}), and (\textit{c}), respectively. The shaded grey regions denote the envelope of variance statistics across all cases in figure~\ref{fig:4}.}
  \label{fig:15}
\end{figure}

Figure~\ref{fig:15} presents a domain-size sensitivity test comparing variance statistics for the moderate-convection case c6 ($-z_i/L = 40$) simulated on its original computational domain ($L_x = L_y = 12\,\mathrm{km} \approx 10.9 z_i$) and the smaller domain ($L_x = L_y = 5.76\,\mathrm{km} \approx 5.2 z_i$) used for the more convective cases in table~\ref{tab:cases}. The simulation conditions are identical aside from the domain size. The normalised spanwise (\textit{b}) and vertical (\textit{c}) velocity variance profiles exhibit close agreement across the entire boundary layer depth, with the mixed-layer peak in $w'^{2\star}$ increasing by less than $4\%$ in the smaller domain.  
There is a modest decrease in streamwise variance (\textit{a}), but the decrease is small relative to the differences across the convective cases.  
This quantitative alignment demonstrates that the energy is sufficiently contained within the smaller computational domain for the more convective cases when the longitudinal rolls have fragmented, and the domain size does not impact the conclusions of the study. 

\section{SGS variance sensitivity}
\label{sec:appB}

\begin{figure}
  \centering
  \includegraphics[width=1.0\textwidth]{figures/global_legend.eps} 
  \vspace{0.2cm}
  \includegraphics[width=1.0\textwidth]{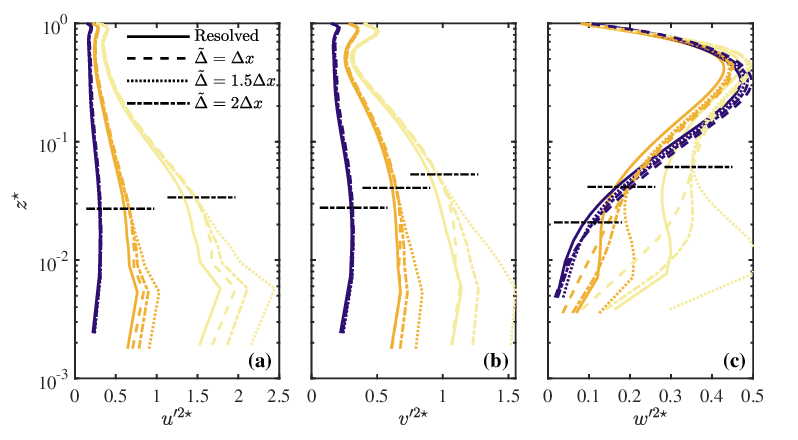}
  \caption{Resolved variances (solid lines) in cases c1, c4, and c12 compared with total variances using a range of filter scales to estimate the SGS contribution -- $\tilde{\rmDelta} = \rmDelta x$ (dashed lines), $\tilde{\rmDelta} = 1.5\rmDelta x$ (dotted lines), and $\tilde{\rmDelta} = 2\rmDelta x$ (dash-dotted lines) -- for the normalised (\textit{a}) streamwise, (\textit{b}) spanwise, and (\textit{c}) vertical velocity components. The near-wall region below the horizontal cutoffs is excluded from the main analysis.}
  \label{fig:16}
\end{figure}

The LES grid filter is assumed to lie within the inertial subrange when the SGS variance is estimated, so that the unresolved energy can be extrapolated from the resolved scales using a $k^{-5/3}$ spectral slope \citep{Knaepen2002,Salesky2012,Hossen2021}. The extrapolation is based on the Leonard stress, which represents the energy difference between the primary grid filter and a coarser test filter.

This assumption becomes increasingly questionable close to the wall because the characteristic turbulent length scales are constrained by the boundary. The integral scale can then approach the grid scale, leaving insufficient separation between the energy-containing range and the nominal inertial subrange. To quantify this uncertainty, the SGS variance was estimated using two primary grid-filter widths, $\tilde{\rmDelta}=\rmDelta x$ and $\tilde{\rmDelta}=1.5\rmDelta x$, where $\rmDelta x$ is the horizontal grid spacing. A test filter twice as coarse was applied in each case.

Figure~\ref{fig:16} illustrates the sensitivity of the estimated total velocity variances to the filter scale used to determine the SGS contribution. A third filter using $\tilde{\rmDelta}=2\rmDelta x$ is shown in the figure to illustrate the strong scale dependence close to the wall, but is not used in determining the cutoff.

The SGS estimate is considered reliable where the resulting total variance is insensitive to the choice of grid filter. In the present analysis, the cutoff height is therefore defined as the lowest height at which the total variances obtained with $\rmDelta x$ and $1.5\rmDelta x$ differ by less than $1\%$. Only data above this height are retained for the surface-layer fits and the variance analyses presented in the main text. This criterion ensures that the reported trends are based on the region where $\rmDelta x$ is well within the inertial subrange and the SGS variance extrapolation is accurate. 

\end{appen}

\bibliographystyle{jfm}
\bibliography{jfm}

@PhdThesis{Albertson1996,
  author  = {J. D. Albertson},
  title   = {Large eddy simulation of land-atmosphere interaction},
  school  = {University of California, Davis},
  year    = {1996}
}

@Article{Albertson1999,
  author  = {J. D. Albertson and M. B. Parlange},
  title   = {Surface length scales and shear stress: Implications for land-atmosphere interaction over complex terrain},
  journal = {Water Resour. Res.},
  volume  = {35},
  pages   = {2121--2132},
  year    = {1999},
  doi     = {10.1029/1999WR900094}
}

@Article{Andren1994,
  author  = {A. Andr{\'e}n and A. R. Brown and P. J. Mason and J. Graf and U. Schumann and C.-H. Moeng and F. T. M. Nieuwstadt},
  title   = {Large-eddy simulation of a neutrally stratified boundary layer: A comparison of four computer codes},
  journal = {Q. J. R. Meteorol. Soc.},
  volume  = {120},
  pages   = {1457--1484},
  year    = {1994},
  doi     = {10.1002/qj.49712052003}
}

@Article{Baars2020,
  author  = {W. J. Baars and I. Marusic},
  title   = {Data-driven decomposition of the streamwise turbulence kinetic energy in boundary layers. {P}art 2: {I}ntegrated energy and {$A_1$}},
  journal = {J. Fluid Mech.},
  volume  = {882},
  pages   = {A26},
  year    = {2020},
  doi     = {10.1017/jfm.2019.835}
}

@Article{Baidya2021,
  author  = {R. Baidya and J. Philip and N. Hutchins and J. P. Monty and I. Marusic},
  title   = {Spanwise velocity statistics in high-{R}eynolds-number turbulent boundary layers},
  journal = {J. Fluid Mech.},
  volume  = {913},
  pages   = {A35},
  year    = {2021},
  doi     = {10.1017/jfm.2020.1129}
}

@Article{Banerjee2013,
  author  = {T. Banerjee and G. G. Katul},
  title   = {Logarithmic scaling in the longitudinal velocity variance explained by a spectral budget},
  journal = {Phys. Fluids},
  volume  = {25},
  number  = {12},
  pages   = {125106},
  year    = {2013},
  doi     = {10.1063/1.4837876}
}

@Article{Banerjee2015,
  author  = {T. Banerjee and G. G. Katul and S. T. Salesky and M. Chamecki},
  title   = {Revisiting the formulations for the longitudinal velocity variance in the unstable atmospheric surface layer},
  journal = {Q. J. R. Meteorol. Soc.},
  volume  = {141},
  pages   = {1699--1711},
  year    = {2015},
  doi     = {10.1002/qj.2472}
}

@Article{Bernardini2014,
  author  = {M. Bernardini and S. Pirozzoli and P. Orlandi},
  title   = {Velocity statistics in turbulent channel flow up to {$Re_\tau = 4000$}},
  journal = {J. Fluid Mech.},
  volume  = {742},
  pages   = {171--191},
  year    = {2014},
  doi     = {10.1017/jfm.2013.674}
}

@Article{Blass2020,
  author  = {A. Blass and X. Zhu and R. Verzicco and D. Lohse and R. J. A. M. Stevens},
  title   = {Flow organization and heat transfer in turbulent wall sheared thermal convection},
  journal = {J. Fluid Mech.},
  volume  = {897},
  pages   = {A22},
  year    = {2020},
  doi     = {10.1017/jfm.2020.378}
}

@Article{BouZeid2005,
  author  = {E. Bou-Zeid and C. Meneveau and M. B. Parlange},
  title   = {A scale-dependent {L}agrangian dynamic model for large eddy simulation of complex turbulent flows},
  journal = {Phys. Fluids},
  volume  = {17},
  pages   = {025105},
  year    = {2005},
  doi     = {10.1063/1.1839152}
}

@Article{BouZeid2018,
  author  = {E. Bou-Zeid and X. Gao and C. Ansorge and G. G. Katul},
  title   = {On the role of return to isotropy in wall-bounded turbulent flows with buoyancy},
  journal = {J. Fluid Mech.},
  volume  = {856},
  pages   = {61--78},
  year    = {2018},
  doi     = {10.1017/jfm.2018.693}
}

@Book{Brutsaert1982,
  author    = {W. Brutsaert},
  title     = {Evaporation into the Atmosphere: Theory, History, and Applications},
  publisher = {Springer},
  year      = {1982},
  doi       = {10.1007/978-94-017-1497-6}
}

@Article{Caughey1979,
  author  = {S. J. Caughey and S. G. Palmer},
  title   = {Some aspects of turbulence structure through the depth of the convective boundary layer},
  journal = {Q. J. R. Meteorol. Soc.},
  volume  = {105},
  pages   = {811--827},
  year    = {1979},
  doi     = {10.1002/qj.49710544606}
}

@Article{ChavezMedina2025,
  author  = {V. {Ch{\'a}vez-Medina} and J. P. Mellado and M. Wilczek},
  title   = {Statistical analysis of vertical velocity and buoyancy in convective boundary layers},
  journal = {J. Fluid Mech.},
  volume  = {1011},
  pages   = {A40},
  year    = {2025},
  doi     = {10.1017/jfm.2025.286}
}

@Article{Deardorff1970,
  author  = {J. W. Deardorff},
  title   = {Convective velocity and temperature scales for the unstable planetary boundary layer and for {R}ayleigh convection},
  journal = {J. Atmos. Sci.},
  volume  = {27},
  pages   = {1211--1213},
  year    = {1970},
  doi     = {10.1175/1520-0469(1970)027<1211:CVATSF>2.0.CO;2}
}

@Article{deRoode2022,
  author  = {S. R. {de Roode} and A. P. Siebesma and F. Jansson and M. Janssens},
  title   = {Dependency of mesoscale organization on grid anisotropy in large-eddy simulations of convective boundary layers at {Gray Zone} resolutions},
  journal = {J. Adv. Model. Earth Syst.},
  volume  = {14},
  pages   = {e2022MS003095},
  year    = {2022},
  doi     = {10.1029/2022MS003095}
}

@Article{Dewani2023,
  author  = {N. Dewani and M. Sakradzija and L. Schlemmer and R. Leinweber and J. Schmidli},
  title   = {Dependency of vertical velocity variance on meteorological conditions in the convective boundary layer},
  journal = {Atmos. Chem. Phys.},
  volume  = {23},
  pages   = {4045--4058},
  year    = {2023},
  doi     = {10.5194/acp-23-4045-2023}
}

@Article{Ding2018,
  author  = {M. Ding and K. X. Nguyen and S. Liu and M. J. Otte and C. Tong},
  title   = {Investigation of the pressure--strain-rate correlation and pressure fluctuations in convective and near-neutral atmospheric surface layers},
  journal = {J. Fluid Mech.},
  volume  = {854},
  pages   = {88--120},
  year    = {2018},
  doi     = {10.1017/jfm.2018.576}
}

@Article{Foken2006,
  author  = {T. Foken},
  title   = {50 years of the {M}onin--{O}bukhov similarity theory},
  journal = {Boundary-Layer Meteorol.},
  volume  = {119},
  pages   = {431--447},
  year    = {2006},
  doi     = {10.1007/s10546-006-9048-6}
}

@Book{Garratt1992,
  author    = {J. R. Garratt},
  title     = {The Atmospheric Boundary Layer},
  publisher = {Cambridge University Press},
  year      = {1992}
}

@Article{Glendening2000,
  author  = {J. W. Glendening},
  title   = {Budgets of Lineal and Nonlineal Turbulent Kinetic Energy under Strong Shear Conditions},
  journal = {J. Atmos. Sci.},
  volume  = {57},
  pages   = {2297--2318},
  year    = {2000},
  doi     = {10.1175/1520-0469(2000)057<2297:BOLANT>2.0.CO;2}
}

@Article{Greene2025,
  author  = {B. R. Greene and L. M. Otterstatter and S. T. Salesky},
  title   = {How representative are uncrewed aircraft system measurements of the convective boundary layer?},
  journal = {Geophys. Res. Lett.},
  volume  = {52},
  pages   = {e2024GL114291},
  year    = {2025},
  doi     = {10.1029/2024GL114291}
}

@Article{Grossman1982,
  author  = {R. L. Grossman},
  title   = {An analysis of vertical velocity spectra obtained in the {BOMEX} fair-weather, trade-wind boundary layer},
  journal = {Boundary-Layer Meteorol.},
  volume  = {23},
  pages   = {323--357},
  year    = {1982},
  doi     = {10.1007/BF00121120}
}

@Article{Heinze2016,
  author  = {R. Heinze and D. Mironov and S. Raasch},
  title   = {Analysis of pressure-strain and pressure gradient-scalar covariances in cloud-topped boundary layers: A large-eddy simulation study},
  journal = {J. Adv. Model. Earth Syst.},
  volume  = {8},
  pages   = {3--30},
  year    = {2016},
  doi     = {10.1002/2015MS000508}
}

@Article{Heisel2024,
  author  = {M. Heisel and M. Chamecki},
  title   = {On the departure from {M}onin--{O}bukhov surface similarity and transition to the convective mixed layer},
  journal = {Boundary-Layer Meteorol.},
  volume  = {190},
  pages   = {28},
  year    = {2024},
  doi     = {10.1007/s10546-024-00870-0}
}

@Article{Heisel2026,
  author  = {M. Heisel and R. Deshpande and G. G. Katul},
  title   = {On the wall-normal velocity variance in canonical wall-bounded turbulence},
  journal = {J. Fluid Mech.},
  volume  = {1033},
  pages   = {A30},
  year    = {2026},
  doi     = {10.1017/jfm.2026.11450}
}

@Article{Hogstrom1990,
  author  = {U. H{\"o}gstr{\"o}m},
  title   = {Analysis of Turbulence Structure in the Surface Layer with a Modified Similarity Formulation for Near Neutral Conditions},
  journal = {J. Atmos. Sci.},
  volume  = {47},
  pages   = {1949--1972},
  year    = {1990},
  doi     = {10.1175/1520-0469(1990)047<1949:AOTSIT>2.0.CO;2}
}

@Article{Hojstrup1982,
  author  = {J. H{\o}jstrup},
  title   = {Velocity Spectra in the Unstable Planetary Boundary Layer},
  journal = {J. Atmos. Sci.},
  volume  = {39},
  pages   = {2239--2248},
  year    = {1982},
  doi     = {10.1175/1520-0469(1982)039<2239:VSITUP>2.0.CO;2}
}

@Article{Hossen2021,
  author  = {M. K. Hossen and A. {Mulayath Variyath} and J. M. Alam},
  title   = {Statistical Analysis of Dynamic Subgrid Modeling Approaches in Large Eddy Simulation},
  journal = {Aerospace},
  volume  = {8},
  number  = {12},
  pages   = {375},
  year    = {2021},
  doi     = {10.3390/aerospace8120375}
}

@Article{Huang2022,
  author  = {K. Y. Huang and G. G. Katul},
  title   = {Profiles of high-order moments of longitudinal velocity explained by the random sweeping decorrelation hypothesis},
  journal = {Phys. Rev. Fluids},
  volume  = {7},
  pages   = {044603},
  year    = {2022},
  doi     = {10.1103/PhysRevFluids.7.044603}
}

@Article{Huang2026,
  author  = {J. Huang and M. Chamecki and Q. Li and B. Chen},
  title   = {The role of longitudinal alignment between surface and bottom forcing on the full-column turbulence mixing in the coastal ocean},
  journal = {Ocean Model.},
  volume  = {199},
  pages   = {102637},
  year    = {2026},
  doi     = {10.1016/j.ocemod.2025.102637}
}

@Article{Hunt2000,
  author  = {J. C. R. Hunt and J. F. Morrison},
  title   = {Eddy structure in turbulent boundary layers},
  journal = {Eur. J. Mech. B/Fluids},
  volume  = {19},
  pages   = {673--694},
  year    = {2000},
  doi     = {10.1016/S0997-7546(00)00129-1}
}

@Article{Jayaraman2021,
  author  = {B. Jayaraman and J. G. Brasseur},
  title   = {Transition in atmospheric boundary layer turbulence structure from neutral to convective, and large-scale rolls},
  journal = {J. Fluid Mech.},
  volume  = {913},
  pages   = {A42},
  year    = {2021},
  doi     = {10.1017/jfm.2021.3}
}

@Article{Kaimal1976,
  author  = {J. C. Kaimal and J. C. Wyngaard and D. A. Haugen and O. R. Cot\'e and Y. Izumi and S. J. Caughey and C. J. Readings},
  title   = {Turbulence Structure in the Convective Boundary Layer},
  journal = {J. Atmos. Sci.},
  volume  = {33},
  pages   = {2152--2169},
  year    = {1976},
  doi     = {10.1175/1520-0469(1976)033<2152:TSITCB>2.0.CO;2}
}

@Book{Kaimal1994,
  author    = {J. C. Kaimal and J. J. Finnigan},
  title     = {Atmospheric Boundary Layer Flows: Their Structure and Measurement},
  publisher = {Oxford University Press},
  year      = {1994},
  doi       = {10.1093/oso/9780195062397.001.0001}
}

@Article{Khanna1997,
  author  = {S. Khanna and J. G. Brasseur},
  title   = {Analysis of {M}onin--{O}bukhov similarity from large-eddy simulation},
  journal = {J. Fluid Mech.},
  volume  = {345},
  pages   = {251--286},
  year    = {1997},
  doi     = {10.1017/S0022112097006277}
}

@Article{Khanna1998,
  author  = {S. Khanna and J. G. Brasseur},
  title   = {Three-dimensional buoyancy- and shear-induced local structure of the atmospheric boundary layer},
  journal = {J. Atmos. Sci.},
  volume  = {55},
  pages   = {710--743},
  year    = {1998},
  doi     = {10.1175/1520-0469(1998)055<0710:TDBASI>2.0.CO;2}
}

@Article{Knaepen2002,
  author  = {B. Knaepen and O. Debliquy and D. Carati},
  title   = {Subgrid-scale energy and pseudo pressure in large-eddy simulation},
  journal = {Phys. Fluids},
  volume  = {14},
  pages   = {4235--4241},
  year    = {2002},
  doi     = {10.1063/1.1514643}
}

@Article{Kumar2006,
  author  = {V. Kumar and J. Kleissl and C. Meneveau and M. B. Parlange},
  title   = {Large-eddy simulation of a diurnal cycle of the atmospheric boundary layer: Atmospheric stability and scaling issues},
  journal = {Water Resour. Res.},
  volume  = {42},
  pages   = {W06D09},
  year    = {2006},
  doi     = {10.1029/2005WR004651}
}

@Article{Kunkel2006,
  author  = {G. J. Kunkel and I. Marusic},
  title   = {Study of the near-wall-turbulent region of the high-{Reynolds}-number boundary layer using an atmospheric flow},
  journal = {J. Fluid Mech.},
  volume  = {548},
  pages   = {375--402},
  year    = {2006},
  doi     = {10.1017/S0022112005007780}
}

@Article{Lee2015,
  author  = {M. Lee and R. D. Moser},
  title   = {Direct numerical simulation of turbulent channel flow up to {$Re_\tau \approx 5200$}},
  journal = {J. Fluid Mech.},
  volume  = {774},
  pages   = {395--415},
  year    = {2015},
  doi     = {10.1017/jfm.2015.268}
}

@Article{Lee2019,
  author  = {M. Lee and R. D. Moser},
  title   = {Spectral analysis of the budget equation in turbulent channel flows at high {R}eynolds number},
  journal = {J. Fluid Mech.},
  volume  = {860},
  pages   = {886--938},
  year    = {2019},
  doi     = {10.1017/jfm.2018.903}
}

@Article{LeMone1976,
  author  = {M. A. LeMone},
  title   = {Modulation of turbulence energy by longitudinal rolls in an unstable planetary boundary layer},
  journal = {J. Atmos. Sci.},
  volume  = {33},
  pages   = {1308--1320},
  year    = {1976},
  doi     = {10.1175/1520-0469(1976)033<1308:MOTEBL>2.0.CO;2}
}

@Article{Lenschow1980,
  author  = {D. H. Lenschow and J. C. Wyngaard and W. T. Pennell},
  title   = {Mean-field and second-moment budgets in a baroclinic, convective boundary layer},
  journal = {J. Atmos. Sci.},
  volume  = {37},
  pages   = {1313--1326},
  year    = {1980},
  doi     = {10.1175/1520-0469(1980)037<1313:MFASMB>2.0.CO;2}
}

@Article{Lenschow2012,
  author  = {D. H. Lenschow and M. Lothon and S. D. Mayor and P. P. Sullivan and G. Canut},
  title   = {A comparison of higher-order vertical velocity moments in the convective boundary layer from lidar with in situ measurements and large-eddy simulation},
  journal = {Boundary-Layer Meteorol.},
  volume  = {143},
  number  = {1},
  pages   = {107--123},
  year    = {2012},
  doi     = {10.1007/s10546-011-9615-3}
}

@book{Lumley1964,
  author    = {Lumley, J. L. and Panofsky, H. A.},
  title     = {The Structure of Atmospheric Turbulence},
  series    = {Interscience Monographs and Texts in Physics and Astronomy},
  volume    = {12},
  publisher = {Interscience Publishers},
  address   = {New York},
  year      = {1964}
}

@Article{Maronga2017,
  author  = {B. Maronga and J. Reuder},
  title   = {On the formulation and universality of {M}onin--{O}bukhov similarity functions for mean gradients and standard deviations in the unstable surface layer: {R}esults from surface-layer-resolving large-eddy simulations},
  journal = {J. Atmos. Sci.},
  volume  = {74},
  pages   = {989--1010},
  year    = {2017},
  doi     = {10.1175/JAS-D-16-0186.1}
}

@Article{Marusic2013,
  author  = {I. Marusic and J. P. Monty and M. Hultmark and A. J. Smits},
  title   = {On the logarithmic region in wall turbulence},
  journal = {J. Fluid Mech.},
  volume  = {716},
  pages   = {R3},
  year    = {2013},
  doi     = {10.1017/jfm.2012.511}
}

@Article{Marusic2019,
  author  = {I. Marusic and J. P. Monty},
  title   = {Attached Eddy Model of Wall Turbulence},
  journal = {Annu. Rev. Fluid Mech.},
  volume  = {51},
  pages   = {49--74},
  year    = {2019},
  doi     = {10.1146/annurev-fluid-010518-040427}
}

@Article{Mironov2000,
  author  = {D. V. Mironov and V. M. Gryanik and C.-H. Moeng and D. J. Olbers and T. H. Warncke},
  title   = {Vertical turbulence structure and second-moment budgets in convection with rotation: A large-eddy simulation study},
  journal = {Q. J. R. Meteorol. Soc.},
  volume  = {126},
  number  = {563},
  pages   = {477--515},
  year    = {2000},
  doi     = {10.1002/qj.49712656306}
}

@Article{Moeng1984,
  author  = {C.-H. Moeng},
  title   = {A large-eddy-simulation model for the study of planetary boundary-layer turbulence},
  journal = {J. Atmos. Sci.},
  volume  = {41},
  pages   = {2052--2062},
  year    = {1984},
  doi     = {10.1175/1520-0469(1984)041<2052:ALESMF>2.0.CO;2}
}

@Article{Moeng1986,
  author  = {C.-H. Moeng and J. C. Wyngaard},
  title   = {An analysis of closures for pressure-scalar covariances in the convective boundary layer},
  journal = {J. Atmos. Sci.},
  volume  = {43},
  pages   = {2499--2513},
  year    = {1986},
  doi     = {10.1175/1520-0469(1986)043<2499:AAOCFP>2.0.CO;2}
}

@Article{Moeng1994,
  author  = {C.-H. Moeng and P. P. Sullivan},
  title   = {A comparison of shear- and buoyancy-driven planetary boundary layer flows},
  journal = {J. Atmos. Sci.},
  volume  = {51},
  pages   = {999--1022},
  year    = {1994},
  doi     = {10.1175/1520-0469(1994)051<0999:ACOSAB>2.0.CO;2}
}

@Article{Moraes2005,
  author  = {O. L. L. Moraes and O. C. Acevedo and G. A. Degrazia and D. Anfossi and R. {da Silva} and V. Anabor},
  title   = {Surface layer turbulence parameters over a complex terrain},
  journal = {Atmos. Environ.},
  volume  = {39},
  pages   = {3103--3112},
  year    = {2005},
  doi     = {10.1016/j.atmosenv.2005.01.046}
}

@Article{Nguyen2013,
  author  = {K. X. Nguyen and T. W. Horst and S. P. Oncley and C. Tong},
  title   = {Measurements of the budgets of the subgrid-scale stress and temperature flux in a convective atmospheric surface layer},
  journal = {J. Fluid Mech.},
  volume  = {729},
  pages   = {388--422},
  year    = {2013},
  doi     = {10.1017/jfm.2013.302}
}

@Article{Nieuwstadt1984,
  author  = {F. T. M. Nieuwstadt},
  title   = {The turbulent structure of the stable, nocturnal boundary layer},
  journal = {J. Atmos. Sci.},
  volume  = {41},
  pages   = {2202--2216},
  doi     = {10.1175/1520-0469(1984)041<2202:TTSOTS>2.0.CO;2},
  year    = {1984}
}

@incollection{Nieuwstadt1993,
  author    = {Nieuwstadt, F. T. M. and Mason, P. J. and Moeng, C.-H. and Schumann, U.},
  title     = {Large-eddy simulation of the convective boundary layer: {A} comparison of four computer codes},
  booktitle = {Turbulent Shear Flows 8},
  editor    = {Durst, F. and Friedrich, R. and Launder, B. E. and Schmidt, F. W. and Schumann, U. and Whitelaw, J. H.},
  pages     = {343--367},
  publisher = {Springer},
  address   = {Berlin, Heidelberg},
  year      = {1993},
  doi       = {10.1007/978-3-642-77674-8_24}
}

@Article{Pandey2018,
  author  = {A. Pandey and J. D. Scheel and J. Schumacher},
  title   = {Turbulent superstructures in {R}ayleigh--{B}\'enard convection},
  journal = {Nat. Commun.},
  volume  = {9},
  pages   = {2118},
  year    = {2018},
  doi     = {10.1038/s41467-018-04478-0}
}

@Article{Panofsky1977,
  author  = {H. A. Panofsky and H. Tennekes and D. H. Lenschow and J. C. Wyngaard},
  title   = {The characteristics of turbulent velocity components in the surface layer under convective conditions},
  journal = {Boundary-Layer Meteorol.},
  volume  = {11},
  pages   = {355--361},
  year    = {1977},
  doi     = {10.1007/BF02186086}
}

@Book{Panofsky1984,
  author    = {H. A. Panofsky and J. A. Dutton},
  title     = {Atmospheric Turbulence: Models and Methods for Engineering Applications},
  publisher = {Wiley},
  year      = {1984}
}

@Article{Perry1987,
  author  = {A. E. Perry and K. L. Lim and S. M. Henbest},
  title   = {An experimental study of the turbulence structure in smooth- and rough-wall boundary layers},
  journal = {J. Fluid Mech.},
  volume  = {177},
  pages   = {437--466},
  year    = {1987},
  doi     = {10.1017/S0022112087001034}
}

@Article{Perry1990,
  author  = {A. E. Perry and J. D. Li},
  title   = {Experimental support for the attached-eddy hypothesis in zero-pressure-gradient turbulent boundary layers},
  journal = {J. Fluid Mech.},
  volume  = {218},
  pages   = {405--438},
  year    = {1990},
  doi     = {10.1017/S0022112090001057}
}

@Article{Pirozzoli2024,
  author  = {S. Pirozzoli},
  title   = {On the streamwise velocity variance in the near-wall region of turbulent flows},
  journal = {J. Fluid Mech.},
  volume  = {989},
  pages   = {A5},
  year    = {2024},
  doi     = {10.1017/jfm.2024.467}
}

@Book{Pope2000,
  author    = {S. B. Pope},
  title     = {Turbulent Flows},
  publisher = {Cambridge University Press},
  year      = {2000},
  doi       = {10.1017/CBO9780511840531}
}

@Article{PorteAgel2000,
  author  = {F. Port\'e-Agel and C. Meneveau and M. B. Parlange},
  title   = {A scale-dependent dynamic model for large-eddy simulation: application to a neutral atmospheric boundary layer},
  journal = {J. Fluid Mech.},
  volume  = {415},
  pages   = {261--284},
  year    = {2000},
  doi     = {10.1017/S0022112000008776}
}

@Article{Priestley1954,
  author  = {C. H. B. Priestley},
  title   = {Convection from a large horizontal surface},
  journal = {Aust. J. Phys.},
  volume  = {7},
  pages   = {176--201},
  year    = {1954},
  doi     = {10.1071/PH540176}
}

@Article{Qin2025,
  author  = {Y. Qin and G. G. Katul and H. Liu and D. Li},
  title   = {Asymptotic coefficients of the attached-eddy model derived from an adiabatic atmosphere},
  journal = {J. Fluid Mech.},
  volume  = {1011},
  pages   = {A29},
  year    = {2025},
  doi     = {10.1017/jfm.2025.394}
}

@Article{Salesky2012,
  author  = {S. T. Salesky and M. Chamecki},
  title   = {A similarity model of subfilter-scale energy for large-eddy simulations of the atmospheric boundary layer},
  journal = {Boundary-Layer Meteorol.},
  volume  = {145},
  pages   = {69--91},
  year    = {2012},
  doi     = {10.1007/s10546-011-9618-0}
}

@Article{Salesky2017,
  author  = {S. T. Salesky and M. Chamecki and E. Bou-Zeid},
  title   = {On the nature of the transition between roll and cellular organization in the convective boundary layer},
  journal = {Boundary-Layer Meteorol.},
  volume  = {163},
  pages   = {41--68},
  year    = {2017},
  doi     = {10.1007/s10546-016-0220-3}
}

@Article{Salesky2018,
  author  = {S. T. Salesky and W. Anderson},
  title   = {Buoyancy effects on large-scale motions in convective atmospheric boundary layers: implications for modulation of near-wall processes},
  journal = {J. Fluid Mech.},
  volume  = {856},
  pages   = {135--168},
  year    = {2018},
  doi     = {10.1017/jfm.2018.711}
}

@Article{Sorbjan1986,
  author  = {Z. Sorbjan},
  title   = {On similarity in the atmospheric boundary layer},
  journal = {Boundary-Layer Meteorol.},
  volume  = {34},
  pages   = {377--397},
  year    = {1986},
  doi     = {10.1007/BF00120989}
}

@Article{Sorbjan1988,
  author  = {Z. Sorbjan},
  title   = {Local similarity in the convective boundary layer ({CBL})},
  journal = {Boundary-Layer Meteorol.},
  volume  = {45},
  pages   = {237--250},
  year    = {1988},
  doi     = {10.1007/BF01066672}
}

@Article{Stiperski2023,
  author  = {I. Stiperski and M. Calaf},
  title   = {Generalizing {M}onin--{O}bukhov similarity theory (1954) for complex atmospheric turbulence},
  journal = {Phys. Rev. Lett.},
  volume  = {130},
  pages   = {124001},
  year    = {2023},
  doi     = {10.1103/PhysRevLett.130.124001}
}

@Article{Stoll2020,
  author  = {R. Stoll and J. A. Gibbs and S. T. Salesky and W. Anderson and M. Calaf},
  title   = {Large-Eddy Simulation of the Atmospheric Boundary Layer},
  journal = {Boundary-Layer Meteorol.},
  volume  = {177},
  pages   = {541--581},
  year    = {2020},
  doi     = {10.1007/s10546-020-00556-3}
}

@Book{Stull1988,
  author    = {R. B. Stull},
  title     = {An Introduction to Boundary Layer Meteorology},
  publisher = {Springer Netherlands},
  year      = {1988},
  doi       = {10.1007/978-94-009-3027-8}
}

@Article{Sullivan2011,
  author  = {P. P. Sullivan and E. G. Patton},
  title   = {The effect of mesh resolution on convective boundary layer statistics and structures generated by large-eddy simulation},
  journal = {J. Atmos. Sci.},
  volume  = {68},
  pages   = {2395--2415},
  year    = {2011},
  doi     = {10.1175/JAS-D-10-05010.1}
}

@Article{Sykes1989,
  author  = {R. I. Sykes and D. S. Henn},
  title   = {Large-Eddy Simulation of Turbulent Sheared Convection},
  journal = {J. Atmos. Sci.},
  volume  = {46},
  pages   = {1106--1118},
  year    = {1989},
  doi     = {10.1175/1520-0469(1989)046<1106:LESOTS>2.0.CO;2}
}

@Book{Tennekes1972,
  author    = {H. Tennekes and J. L. Lumley},
  title     = {A First Course in Turbulence},
  publisher = {MIT Press},
  year      = {1972},
  doi       = {10.7551/mitpress/3014.001.0001}
}

@Article{TongNguyen2015,
  author  = {C. Tong and K. X. Nguyen},
  title   = {Multipoint {Monin}--{Obukhov} similarity and its application to turbulence spectra in the convective atmospheric surface layer},
  journal = {J.~Atmos. Sci.},
  year    = {2015},
  volume  = {72},
  pages   = {4337--4348},
  doi     = {10.1175/JAS-D-15-0134.1}
}

@Book{Townsend1976,
  author    = {A. A. Townsend},
  title     = {The Structure of Turbulent Shear Flow},
  publisher = {Cambridge University Press},
  year      = {1976}
}

@Article{Wang2020,
  author  = {Q. Wang and K. L. Chong and R. J. A. M. Stevens and R. Verzicco and D. Lohse},
  title   = {From zonal flow to convection rolls in {R}ayleigh--{B}\'enard convection with free-slip plates},
  journal = {J. Fluid Mech.},
  volume  = {905},
  pages   = {A21},
  year    = {2020},
  doi     = {10.1017/jfm.2020.793}
}

@Article{Wang2021,
  author  = {B. Wang and Q. Wang and Z.-H. Wan and D.-J. Sun},
  title   = {The influence of aspect ratio on flow states in the buoyancy-driven turbulence with free slip boundaries},
  journal = {Int. J. Heat Mass Transfer},
  volume  = {178},
  pages   = {121639},
  year    = {2021},
  doi     = {10.1016/j.ijheatmasstransfer.2021.121639}
}

@Article{Weckwerth1999,
  author  = {T. M. Weckwerth and T. W. Horst and J. W. Wilson},
  title   = {An Observational Study of the Evolution of Horizontal Convective Rolls},
  journal = {Mon. Wea. Rev.},
  volume  = {127},
  pages   = {2160--2179},
  year    = {1999},
  doi     = {10.1175/1520-0493(1999)127<2160:AOSOTE>2.0.CO;2}
}

@Article{Wilczak2001,
  author  = {J. M. Wilczak and S. P. Oncley and S. A. Stage},
  title   = {Sonic Anemometer Tilt Correction Algorithms},
  journal = {Boundary-Layer Meteorol.},
  volume  = {99},
  pages   = {127--150},
  year    = {2001},
  doi     = {10.1023/A:1018966204465}
}

@Article{Wilson2008,
  author  = {J. D. Wilson},
  title   = {{M}onin--{O}bukhov functions for standard deviations of velocity},
  journal = {Boundary-Layer Meteorol.},
  volume  = {129},
  pages   = {353--369},
  year    = {2008},
  doi     = {10.1007/s10546-008-9319-5}
}

@Article{Wyngaard1971,
  author  = {J. C. Wyngaard and O. R. Cot\'e and Y. Izumi},
  title   = {Local free convection, similarity, and the budgets of shear stress and heat flux},
  journal = {J. Atmos. Sci.},
  volume  = {28},
  pages   = {1171--1182},
  year    = {1971},
  doi     = {10.1175/1520-0469(1971)028<1171:LFCSAT>2.0.CO;2}
}

@Article{Wyngaard1974,
  author  = {J. C. Wyngaard and O. R. Cot\'e},
  title   = {The evolution of a convective planetary boundary layer --- A higher-order-closure model study},
  journal = {Boundary-Layer Meteorol.},
  volume  = {7},
  pages   = {289--308},
  year    = {1974},
  doi     = {10.1007/BF00240833}
}

@Book{Wyngaard2010,
  author    = {J. C. Wyngaard},
  title     = {Turbulence in the Atmosphere},
  publisher = {Cambridge University Press},
  year      = {2010},
  doi       = {10.1017/CBO9780511840524}
}

@InCollection{Yaglom1994,
  author    = {A. M. Yaglom},
  title     = {Similarity laws for wall turbulent flows: their limitations and generalizations},
  booktitle = {New Approaches and Concepts in Turbulence},
  editor    = {Th. Dracos and A. Tsinober},
  pages     = {7--27},
  publisher = {Birkh{\"a}user},
  address   = {Basel},
  year      = {1993}
}

@Article{Yang2020,
  author  = {X. I. A. Yang and S. Pirozzoli and M. Abkar},
  title   = {Scaling of velocity fluctuations in statistically unstable boundary-layer flows},
  journal = {J. Fluid Mech.},
  volume  = {886},
  pages   = {A3},
  year    = {2020},
  doi     = {10.1017/jfm.2019.1034}
}

@Article{Yi2025,
  author  = {Y. R. Yi and J. R. Koseff and E. Bou-Zeid},
  title   = {Nonlinear and buoyancy pressure correlations in stably stratified turbulence},
  journal = {J. Fluid Mech.},
  volume  = {1019},
  pages   = {A14},
  year    = {2025},
  doi     = {10.1017/jfm.2025.10605}
}

@Article{Zhou2019,
  author  = {B. Zhou and S. Sun and J. Sun and K. Zhu},
  title   = {The universality of the normalized vertical velocity variance in contrast to the horizontal velocity variance in the convective boundary layer},
  journal = {J. Atmos. Sci.},
  volume  = {76},
  pages   = {1437--1456},
  year    = {2019},
  doi     = {10.1175/JAS-D-18-0325.1}
}

@Article{Zilitinkevich1994,
  author  = {S. Zilitinkevich},
  title   = {A generalized scaling for convective shear flows},
  journal = {Boundary-Layer Meteorol.},
  volume  = {70},
  pages   = {51--78},
  year    = {1994},
  doi     = {10.1007/BF00712523}
}

@Article{Zilitinkevich2006,
  author  = {S. S. Zilitinkevich and J. C. R. Hunt and I. N. Esau and A. A. Grachev and D. P. Lalas and E. Akylas and M. Tombrou and C. W. Fairall and H. J. S. Fernando and A. A. Baklanov and S. M. Joffre},
  title   = {The influence of large convective eddies on the surface-layer turbulence},
  journal = {Q. J. R. Meteorol. Soc.},
  volume  = {132},
  pages   = {1423--1456},
  year    = {2006},
  doi     = {10.1256/qj.05.79}
}

@Article{Zimmerman2019,
  author  = {S. Zimmerman and J. Philip and J. Monty and A. Talamelli and I. Marusic and B. Ganapathisubramani and R. J. Hearst and G. Bellani and R. Baidya and M. Samie and X. Zheng and E. Dogan and L. Mascotelli and J. Klewicki},
  title   = {A comparative study of the velocity and vorticity structure in pipes and boundary layers at friction {R}eynolds numbers up to {$10^4$}},
  journal = {J. Fluid Mech.},
  volume  = {869},
  pages   = {182--213},
  year    = {2019},
  doi     = {10.1017/jfm.2019.182}
}
\end{document}